\pdfoutput=1
\documentclass[twocolumn]{aastex63}
\usepackage{multirow}
\usepackage{booktabs}
\usepackage{amsmath}
\usepackage{threeparttable}
\usepackage{hyperref}			

\hypersetup{colorlinks,
	linkcolor=blue,
	anchorcolor=blue,
	citecolor=blue}
\newcommand{\sersic}{S\'{e}rsic}
\newcommand{\GALFIT}{\texttt{GALFIT}}

\newcommand{\RNum}[1]{\uppercase\expandafter{\romannumeral #1\relax}}

\shorttitle{Morphology of M81}
\shortauthors{Gong et al.}
\graphicspath{{./}{figures/}}

\begin{document}
	
	\title{\textbf{MULTIWAVELENGTH BULGE-DISK DECOMPOSITION FOR THE GALAXY M81 (NGC\,3031).\\
		I. MORPHOLOGY}}
	
	\correspondingauthor{Ye-Wei Mao}
	\email{ywmao@gzhu.edu.cn}
	
	\author{Jun-Yu Gong}
	\affiliation{Center for Astrophysics, GuangZhou University, GuangZhou 510006, People's Republic of China}
	\affiliation{Department of Astronomy, School of Physics and Materials Science, GuangZhou University, GuangZhou 510006, People's Republic of China}
	
	\author{Ye-Wei Mao}
	\affiliation{Center for Astrophysics, GuangZhou University, GuangZhou 510006, People's Republic of China}
	\affiliation{Department of Astronomy, School of Physics and Materials Science, GuangZhou University, GuangZhou 510006, People's Republic of China}
	
	\author{Hua Gao}
	\affiliation{Institute for Astronomy, University of Hawaii, 2680 Woodlawn Drive, Honolulu HI 96822, USA}
	\affiliation{Kavli Institute for the Physics and Mathematics of the Universe (WPI), The University of Tokyo Institutes for Advanced Study, The University of Tokyo, Kashiwa, Chiba 277-8583, Japan}
	
	\author{Si-Yue Yu} \thanks{Humboldt Postdoctoral Fellow}
	\affiliation{Max-Planck-Institut f\"{u}r Radioastronomie, Auf dem H\"{u}gel 69, 53121 Bonn, Germany}

	\begin{center}
		\begin{abstract}
			A panchromatic investigation of morphology for the early-type spiral galaxy M81 is presented in this paper.  We perform bulge-disk decomposition in M81 images at a total of 20 wavebands from FUV to NIR obtained with \emph{GALEX}, \emph{Swift}, SDSS, WIYN, 2MASS, \emph{WISE}, and \emph{Spitzer}.  Morphological parameters such as \sersic{} index, effective radius, position angle, and axis ratio for the bulge and the disk are thus derived at all the wavebands, which enables quantifying the morphological {\it K}-correction for M81 and makes it possible to reproduce images for the bulge and the disk in the galaxy at any waveband.  The morphology as a function of wavelength appears as a variable-slope trend of the \sersic{} index and the effective radius, in which the variations are steep at UV--optical and shallow at optical--NIR bands; the position angle and the axis ratio keep invariable at least at optical--NIR bands.  It is worth noting that, the \sersic{} index for the bulge reaches $\sim 4$--$5$ at optical and NIR bands, but drops to $\sim 1$ at UV bands.  This difference brings forward a caveat that, a classical bulge is likely misidentified for a pseudo-bulge or no bulge at high redshifts where galaxies are observed through rest-frame UV channels with optical telescopes.  The next work of this series is planned to study spatially resolved SEDs for the bulge and the disk, respectively, and thereby explore stellar population properties and star formation/quenching history for the the galaxy composed of the subsystems.
		\end{abstract}
	\end{center}
	
	
	\section{\textbf{INTRODUCTION}}\label{Sec_Intro}
	
	Stellar systems on a galactic scale can be classified into the two fundamental shapes, elliptical/spheroidal and discoidal, according to their morphology.  The two different morphological types originate from different formation mechanisms.  Elliptical/spheroidal systems such as elliptical/spheroidal galaxies or bulges in spiral/lenticular galaxies are considered to be formed by rapid and violent processes \citep[e.g., major mergers,][]{Eggen_1962, Toomre_1977, Dekel_2009, Daniel_2010}; disks in spiral/lenticular galaxies are considered to be formed by slow and gentle processes \citep[e.g., secular evolution,][]{kormendy_1979b, kormendy_2004}.
	
	For early-type spiral/lenticular galaxies, a combination of bulges and disks complicates studies of these galaxies, since measurements of such galaxies as a mixture of bulges and disks would lead to an average effect on results (e.g., \citealt{Mancini_2019}).  Therefore, it is necessary to decompose the galaxies and separate the different galactic structures so as to acquire unbiased knowledge of their nature (\citealt{Debattista_2006, Kendall_2008, Simard_2011}).
	
	Radial profiles of surface brightness are an effective tool for distinguishing between different structures in galaxies.  For both of elliptical/spheroidal and discoidal systems, the surface-brightness profiles can be fitted with the \sersic{} function (\citealt{sersic_1968}). The equation of the \sersic{} function is as follows:
	\begin{equation}\label{eq:sersic}
		\Sigma(r) = \Sigma(e) \exp \left[-\kappa\left(\left(\frac{r}{R_e}\right)^{\frac{1}{n}}-1\right)\right]  ~,
	\end{equation}
	where $\Sigma(r)$ is the surface brightness at the radius $r$, $R_e$ is the effective radius, $\Sigma(e)$ is the surface brightness at the effective radius, $n$ is named as the \sersic{} index , and $\kappa$ is a coefficient correlated with \sersic{} index $n$.  The \sersic{} index $n$ is the parameter distinct between different types of systems.  Empirical statistics have found $n \sim 4$ for elliptical/spheroidal systems (i.e., the de Vaucouleurs law, \citealt{Vaucouleurs_1948}) and $n \sim 1$ for discoidal systems (i.e., the exponential law, \citealt{Freeman_1970}).  The \sersic{} function has been widely applied to the decomposition of galaxies and helped to comprehensively understand formation and evolution of stellar systems on a galactic or subgalactic scale \citep{Peng_2002, Gabor_2009, Peng_2010, Van_2012, Erwin_2015_imfit, 2015_Meert, Gao_2018, Gao_2019}.
	
	At present, most morphological studies of galaxies are conducted at a single waveband (e.g., observed optical wavelength).  In fact, the morphology of galaxies has been found, by modern observations, to vary with wavelength.  The variation of morphology with wavelength, i.e., called the "morphological {\it K}-correction", is mainly ascribed to differences in stellar populations or/and dust attenuation (\citealt{Bohlin_1991, Giavalisco_1996, Mager_2018}).  The morphological {\it K}-correction is likely to induce deviation in the single-band classification, and particularly has impact on studies of high-redshift galaxies for which the number of observable wavebands is limited (\citealt{Kuchinski_2000, Kuchinski_2001, Windhorst_2002, Huertas_2009}).  Even worse, due to the limitations of observation conditions such as the spatial resolution and the signal-to-noise ratio (S/N), there have not been adequate studies that could quantitatively measure multiwavelength morphological parameters, even for nearby galaxies (\citealt{Vulcani_2014, Mager_2018, Psychogyios_2020}).
	
	Aimed at exploring multiwavelength morphology and quantifying the morphological {\it K}-correction, in this paper, we present an investigation of the nearby galaxy M81 (also named NGC\,3031).  M81 is an early-type spiral galaxy (classified as the Hubble type SAab), with the distance $\sim 3.6$ Mpc (\citealt{Madore_1993}) and the angular size $\sim 26\arcmin.9 \times 14\arcmin.1$ (\citealt{Nilson_1973}).  The adjacency of M81 brings on inherently high spatial resolution.  M81 hosts a prominent bulge embedded in the disk.  Space and ground-based telescopes have obtained superb data with nearly full wavelength coverage.  These conditions make M81 to be an ideal target for a multiwavelength morphological study.
	
	The remainder of this paper is structured as follows.  In Section \ref{Sec_Data}, we describe the multiwavelength data, the relevant data processing; Section \ref{Sec_Decomp} presents the methodology for the decomposition; morphological parameters as a function of wavelength are presented in Section \ref{Sec_Result}; in Section \ref{Sec_Disc}, we discuss some questions in the wavelength-dependent morphology and practical applicability of our findings. Conclusions of this work are summarized in Section \ref{Sec_Sum}.  In this paper, we adopt 1 kpc\,arcmin$^{-1}$ at the distance 3.6 Mpc for M81.

	\section{\textbf{MULTIWAVELENGTH DATA AND PRE-PROCESSING}}\label{Sec_Data}
	
	\begin{table*}[!htbp]
		\begin{center}
			\centering
			\caption{Basic Information about the Data Used in This Work}
			\label{table1}
			
			\begin{tabular}{cccccc}
				\toprule{}
				Telescope & Filter & Wavelength & FWHM & Pixel Scale & Reference\\
				& & (\AA) & (arcsec) & (arcsec\,pixel$^{-1}$)\\
				(1) & (2) & (3) & (4) & (5) & (6) \\
				\midrule
				\emph{GALEX} & FUV & 1516 & 4.3 & 1.5 & (a) \\
				\emph{Swift} & UVW2 & 1928 & 2.5 & 1.0 & (b)\\
				\emph{GALEX} & NUV & 2267 & 5.3 & 1.5 & (a) \\
				\emph{Swift} & UVW1 & 2600 & 2.5 & 1.0 & (b)\\
				SDSS & {\it u} & 3551 & 1.3 & 0.396 & (c)\\
				WIYN & B & 4331 & 2.7 & 0.603 & (d)\\
				SDSS & {\it g} & 4686 & 1.3 & 0.396 & (c)\\
				WIYN & V & 5500 & 3.0 & 0.603 & (d)\\
				SDSS & {\it r} & 6166 & 1.3 & 0.396 & (c)\\
				WIYN & R & 6425 & 2.0 & 0.775 & (d)\\
				SDSS & {\it i} & 7480 & 1.3 & 0.396 & (c) \\
				SDSS & {\it z} & 8932 & 1.3 & 0.396 & (c) \\
				2MASS & J & 12000 & 2.0 & 1.0 & (e)\\
				2MASS & H & 16000 & 2.0 & 1.0 & (e) \\
				2MASS & K$_s$ & 22000 & 2.0 & 1.0 & (e)\\
				\emph{WISE} & W1 (3.4 $\mu$m) & 34000 & 6.1 & 1.375 & (f) \\
				\emph{Spitzer}-IRAC & I1 (3.6 $\mu$m) & 36000 & 1.66 & 0.75 & (g) \\
				\emph{Spitzer}-IRAC & I2 (4.5 $\mu$m) & 45000 & 1.72 & 0.75 & (g)\\
				\emph{WISE} & W2 (4.6 $\mu$m) & 46000 & 6.4 & 1.375 & (f)\\
				\emph{Spitzer}-IRAC & I3 (5.8 $\mu$m) & 58000 & 1.88 & 0.75 & (g) \\
				\bottomrule
				
			\end{tabular}
			\begin{tablenotes}
				\footnotesize
				\item \textbf{Notes.} --- Columns:
				(1) the name of the telescope or survey,
				(2) the band/filter name,
				(3) the central wavelength of the filter,
				(4) the full width at the half maximum (FWHM) of the point-spread function (PSF),
				(5) angular scale of the pixel width in the image,
				(6) reference.
				\item \textbf{Reference.}
                (a) \citet{Morrissey_2007}; data are retrieved from \emph{GALEX} Guest Investigator project	(\href{https://asd.gsfc.nasa.gov/archive/galex/}{https://asd.gsfc.nasa.gov/archive/galex/}) and \emph{GALEX} DR6 (\href{http://www.galex.caltech.edu/researcher/data.html}{http://www.galex.caltech.edu/researcher/data.html});
                (b) data are retrieved from SkyView Virtual Observatory (\href{https://skyview.gsfc.nasa.gov/current/cgi/titlepage.pl}{https://skyview.gsfc.nasa.gov/current/cgi/titlepage.pl});
                (c) \citet{Alam_2015}; data are retrieved from SDSS DR12 (\href{https://www.sdss3.org/dr12/}{https://www.sdss3.org/dr12/});
                (d) \citet{cheng_1997, Marcum_2001}; data are retrieved from NASA NED (\href{https://ned.ipac.caltech.edu}{https://ned.ipac.caltech.edu});
                (e) \citet{Skrutskie_2006}; data are retrieved from IRSA (\href{https://irsa.ipac.caltech.edu/frontpage/}{https://irsa.ipac.caltech.edu/frontpage/});
                (f) \citet{Wright_2010}; data are retrieved from IRSA;
                (g) IRAC 3.6 and 4.5 $\mu$m date are retrieved from IRSA, the Spitzer Survey of Stellar Structure in Galaxies \citep[S$^4$G,][]{Sheth_2010, Mu_oz_Mateos_2013, Querejeta_2015}; IRAC-5.8 $\mu$m data are retrieved from IRSA, the Spitzer Infrared Nearby Galaxies Survey \citep[SINGS,][]{Kennicutt_Jr__2003}.
			\end{tablenotes}
		\end{center}
	\end{table*}
	
\begin{figure*}[!htbp]
		\centering
		\plotone{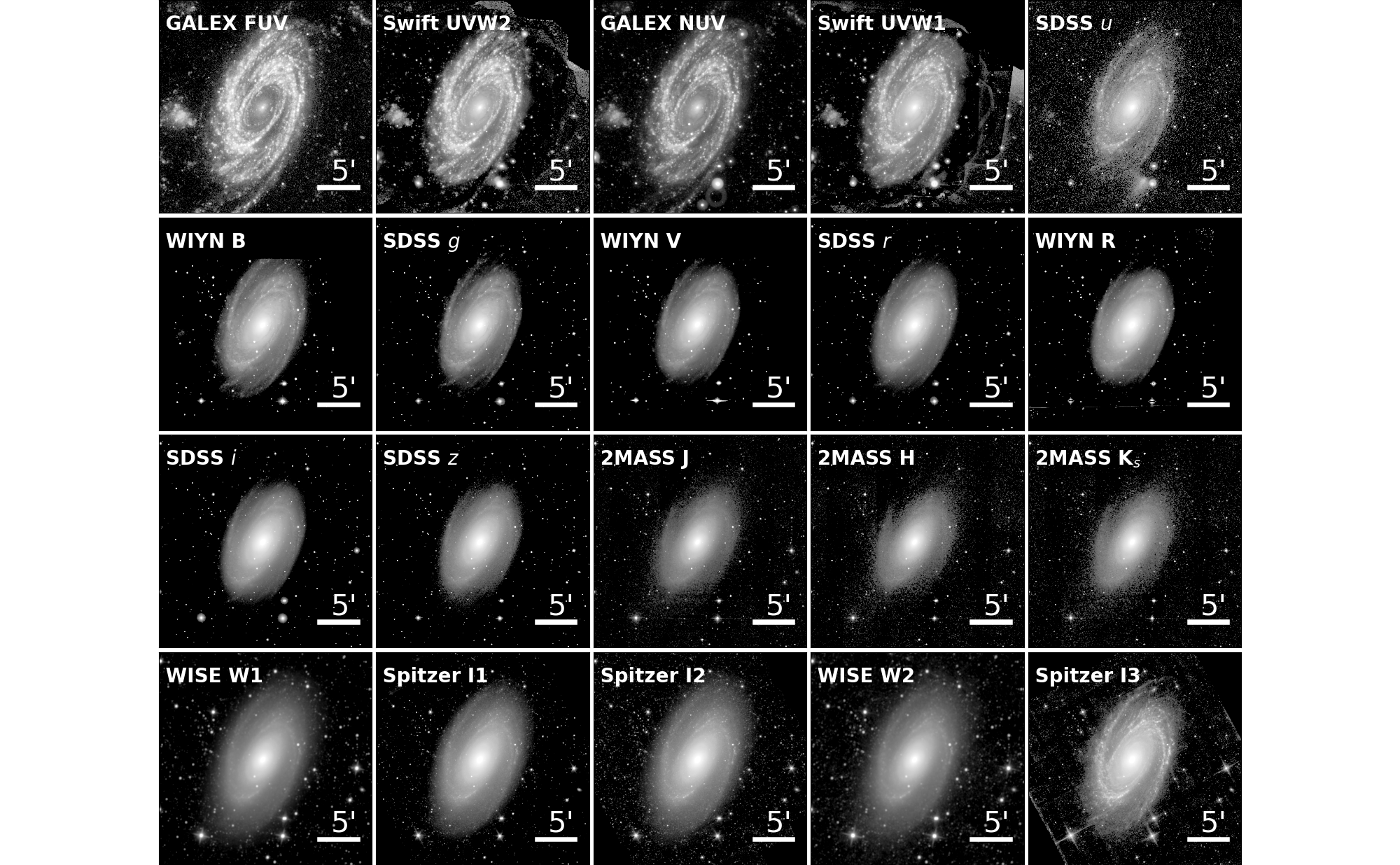}
		\caption{Images at 20 bands for M81 covering the wavelength range from FUV to NIR, obtained with space and ground-based telescopes.  The physical scale is the same in each of the images.  The scale in the bottom-right corner indicates the 5$\arcmin$ length.  North is up and east is to the left.}
		\label{fig:M81pic}
	\end{figure*}

	This work is conducted on the basis of imaging data from FUV to NIR obtained with space and ground-based telescopes/surveys, including \emph{GALEX}, \emph{Swift}, SDSS, WIYN, 2MASS, \emph{WISE}, and \emph{Spitzer}.  All of these data were retrieved from their official websites.  Table \ref{table1} lists basic information about the data used in this work.  Figure \ref{fig:M81pic} shows an atlas of M81 imaged at 20 wavebands from 1516 \AA~to 5.8 $\mu$m.  In this work, correction for the Milky Way (MW) foreground extinction was performed with the \citet{Fitzpatrick_1999} extinction curve, the V-band extinction $A(\mathrm{V})_\mathrm{GAL}$ = 0.220\,mag (\citealt{Schlafly_2011}), and the total-to-selective extinction ratio $R_\mathrm{V}$ = 3.1.
		
	Prior to bulge-disk decomposition of the galaxy, it is necessary to subtract background in each of the images.\footnote{~For \emph{Spitzer} data, although constant background had been officially subtracted before the images were uploaded to the website, we still performed background subtraction for the purpose of treating spatial variations in the background.}  In order to avoid contamination from non-background sources, we at first performed object detection with a $3 \sigma$ threshold and masked them in the images.
	
	The image background was modeled for subtraction through two steps.  The first step is the median-window approach to smoothing the source-masked images at all of the wavebands (\citealt{Mao_2012, Mao_2014}).  In the median-smoothed images, we performed bi-cubic fitting as the second step of the reproduction, during which the background model was interpolated into the masked regions.  The reproduced background was subtracted from the observed image at each waveband.  The uncertainty in the background is defined as the standard deviation in background pixels in each of the background-subtracted images.
	
	\section{\textbf{METHODOLOGY OF THE DECOMPOSITION}}\label{Sec_Decomp}
	
	Bulge-disk decomposition was carried out in the background-subtracted images by using the software \GALFIT{} (\citealt{Peng_2002, Peng_2010}).  Before the decomposition, objects unrelated to the galaxy, such as foreground stars and background galaxies, were identified and masked in the images.  For each-band image, we created a point-spread function (PSF) image from statistics of isolated, unsaturated foreground stars, by fitting with an analytical function \citep[following the method in][]{Anderson_2000, Anderson_2016}.
	
	Sigma images, recording total noise for the observed images, were also created at all of the wavebands.  The construction is formulated as follows (addressed in \citealt{Peng_2002}).
	
	\begin{equation}\label{eq:sigma}
		{\sigma_e^2} = {\sigma_P^2} + {\sigma_R^2} ~,
	\end{equation}
	
	where $\sigma_e$ is the sigma image, $\sigma_P$ is Poisson noise, and $\sigma_R$ is the CCD readout noise.  $\sigma_P$ follows the Poisson distribution.  In order to calculate $\sigma_P$, we converted the units for the original images without background subtraction into electronic counts by multiplying the CCD Gain, and $\sigma_P$ was defined as the square root of the electronic counts.  All of the parameters in Equation (\ref{eq:sigma}) are in units of electronic counts.
	
	Since M81 hosts an AGN at its center (\citealt{Filippenko_1988}), in order to obtain accurate results, we took the AGN into account during the process of the bulge-disk decomposition and hence adopted three components including the bulge, the disk, and the AGN in the model.  The \sersic{} function was used for fitting the bulge and the disk (Equation (\ref{eq:sersic})), while the PSF was used for fitting the AGN.\footnote{~The presence of AGNs will affect morphological estimation of galaxies, if they are not separated from other components in models (\citealt{Povi_2012, Kauffmann_2003, Hopkins_2009}).}  There are six parameters for each \sersic{} component in the model, including the central coordinate ($x$, $y$), the \sersic{} index ($n$), the effective radius ($R_e$), the axis ratio ($q$), the position angle (PA), and the brightness (or magnitude $m_{b}$).  Among these parameters, the central coordinate ($x$, $y$) is set to be constant and equal to the galactic center in all of the images.
	
	Uncertainties in the decomposition are dominated by the background subtraction. Estimation of the uncertainties follows the method in \citet{Huang_2013, Gao_2017}, which empirically estimates influences of the background subtraction on all of the morphological parameters.  For most of the images, we manually add or subtract a constant background value to the background-subtracted images, and then perform the decomposition again in the processed images to obtain new output as the upper or lower limit on the parameters.  The upper and lower limits are taken as the uncertainties in the parameters.  For SDSS-{\it u}, -{\it z}, and 2MASS images with low S/N, this empirical method may yield problematic results such as the \sersic{} index over 20,  because low-brightness sources in these images are easily affected by any change in the background. In order to ensure that the best-fit models after adjusting the background are always physically meaningful, we  fixed the effective radius when calculating the uncertainty of the \sersic{} index, and vice versa.  The uncertainties estimated in this case serve as lower limits only.  This special method is applied only as a backup when the empirical method fails.
	
	\section{\textbf{RESULTS OF THE DECOMPOSITION}}\label{Sec_Result}
	
	\begin{figure*}[!ht]
		\centering
		\plotone{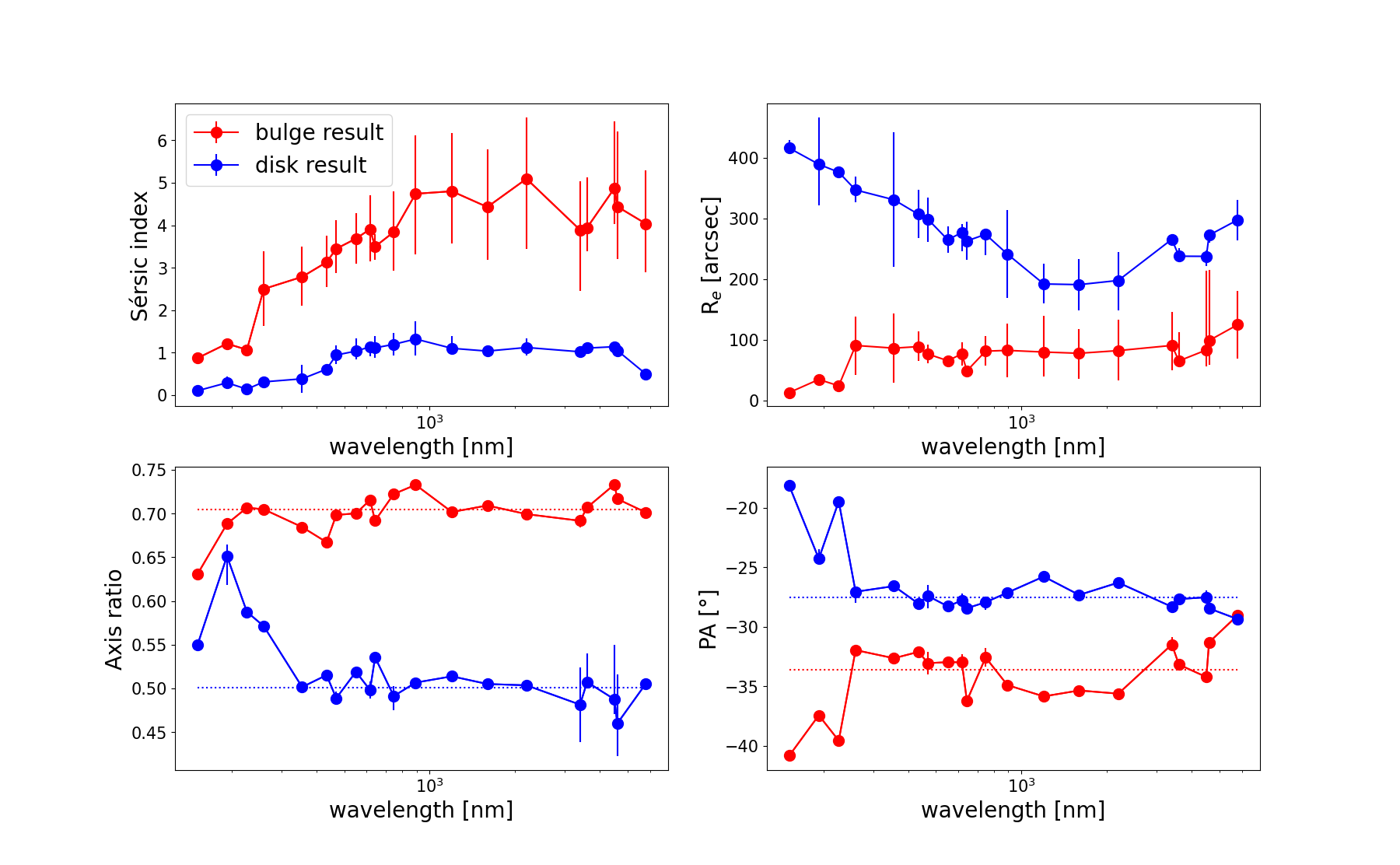}
		\caption{Morphological parameters, including the \sersic{} index (the top-left panel), the effective radius (the top-right panel), the axis ratio (the bottom-left panel), and the position angle (the bottom-right panel), as a function of wavelength for the bulge (red color) and the disk (blue color) in M81.  In each panel, points are connected by solid lines.  In the bottom two panels, dotted lines represent the mean values for the axis ratio (i.e., 0.70 and 0.50 for the bulge and the disk, respectively) and the position angle (i.e., $-33^\circ.62$ and $-27^\circ.52$ for the bulge and the disk, respectively) among the optical and NIR bands.}
		\label{fig:3-component-ser}
	\end{figure*}
	
	\subsection{Morphology as a Function of Wavelength}\label{Sec_Result_WavMor}
	
	\begin{figure*}[!ht]
		\centering
		\plotone{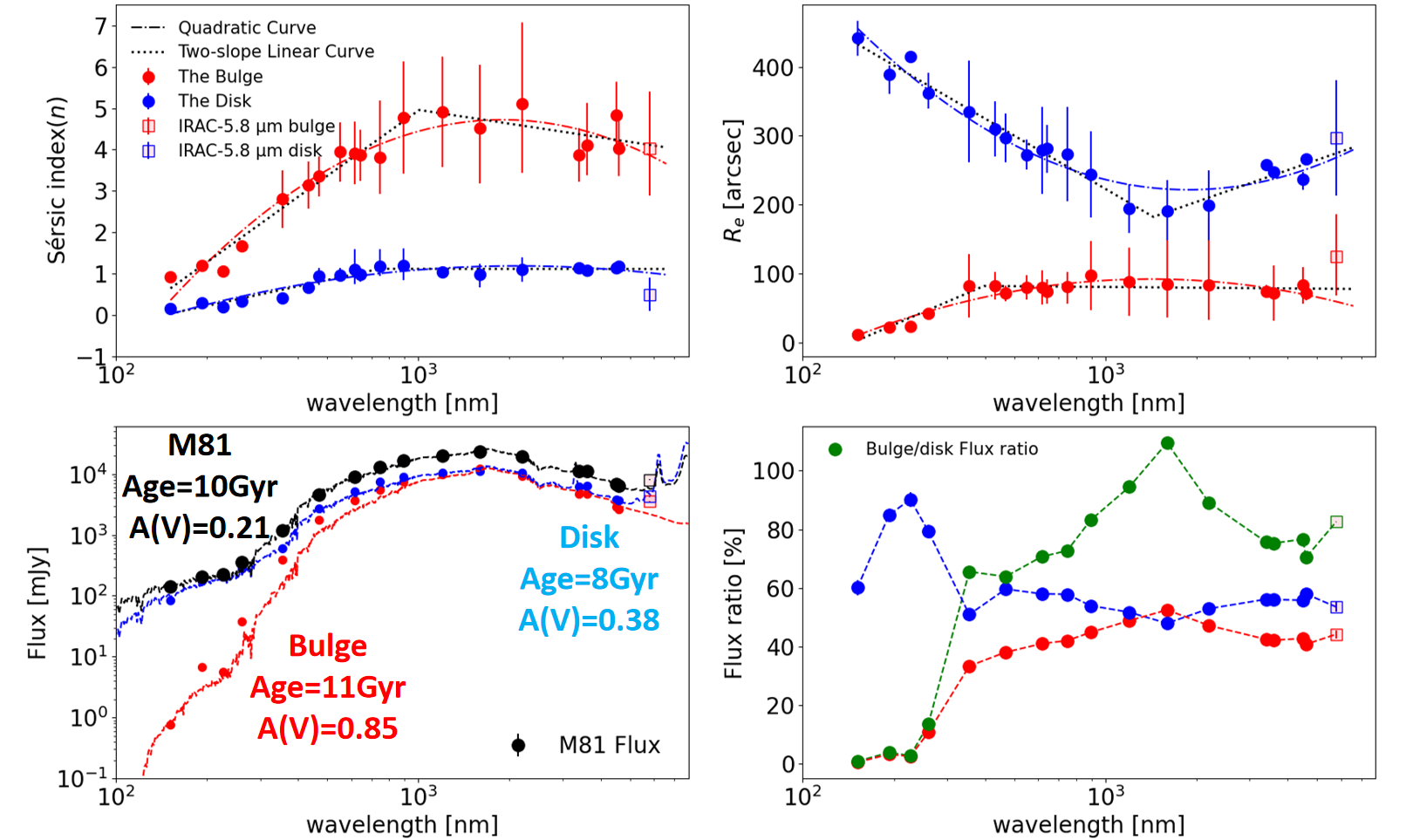}
		\caption{The \sersic{} index (top-left panel), the effective radius (top-right panel), the flux (bottom-left panel), and the flux ratio (bottom-right panel) as a function of wavelength for the bulge (red) and the disk (blue) in M81.  In each panel, the \emph{Spitzer}-IRAC 5.8 $\mu$m band is symbolized as the open square.  In the top panels, dot-dashed and dotted lines represent the best-fit quadratic and piecewise linear curves, respectively, for the wavebands with the \emph{Spitzer}-IRAC 5.8 $\mu$m excluded; in the bottom-left panel, dashed lines are the best-fit model spectra of stellar population synthesis; in the bottom-right panel, dashed lines connect data points.  Black filled circles in the bottom-left panel indicate the total flux of the galaxy.  The flux ratio in the bottom-right panel is defined as the ratio of the bulge (red) or the disk (blue) to the total galaxy enclosed in the same aperture, as well as the ratio of the bulge to the disk (green)}.
		\label{fig:final}
	\end{figure*}
	
	After separately performing the decomposition in each of the images, we obtain morphological parameters including the \sersic{} index $n$, the effective radius $R_e$, the axis ratio $q$, and the position angle PA for the bulge and the disk in M81 at all of the wavebands.  Figure \ref{fig:3-component-ser} shows the variations in the four parameters with wavelength.  We can see that the axis ratio and the position angle are nearly constant except at UV bands.  The deviation in the axis ratio and the position angle is mainly caused by the bias of other substructures such as spiral arms, particularly at the UV bands, as well as low S/N.  In order to reduce the error, we fix $q$ and PA into the mean values for the two parameters in the optical and NIR range.  The decomposition is then performed again to specially obtain variations in the \sersic{} index and the effective radius with wavelength for the bulge and the disk.  Products of the bulge-disk decomposition, including observed, model, residual images, radial profiles of surface brightness, and parameters at the 20 wavebands are presented in detail in Appendix \ref{Sec_Appendix_A}.  The goodness of the fit will be discussed in Section \ref{Sec_Disc}.
	
	With the axis ratio and the position angle fixed, the \sersic{} index and the effective radius as a function of wavelength are displayed in Figure \ref{fig:final}.  We can see that the bulge and the disk present distinct features in the wavelength-dependent morphology.  For the bulge, the \sersic{} index varies remarkably with wavelength from $1516$\,\AA~(i.e., the \emph{GALEX} FUV band) and reaches a peak around $n \sim 5$ at $\lambda \sim 1.2$--$2.2\,\mu$m (i.e., the 2MASS bands), including an intensive increase from $\sim 1$ to $\sim5$ with $\lambda$ from $1516$\,\AA~ to $1.2\,\mu$m, and a moderate decrease from $\sim5$ to $\sim 4$ with $\lambda$ from $2.2\,\mu$m to $4.6\,\mu$m.  The effective radius is almost around $80 \arcsec$ except for the \emph{GALEX} and \emph{Swift} UV-band data which exhibit an increasing trend from 12$\arcsec$ to 43$\arcsec$ with wavelength.  The disk appears with the \sersic{} index slightly increasing from 0.15 to 0.67 in the wavelength range from 1516\,\AA~ (i.e., the \emph{GALEX} FUV band) to 4331\,\AA~ (i.e., the WIYN B band), and nearly constant at $n \sim 1$ from 4686\,\AA~ (i.e., the SDSS {\it g} band) to $4.6\,\mu$m (i.e., the \emph{WISE} W2 band).  The variation in the effective radius emerges with a turning shape similar to that for the bulge but in an opposite direction: including a decrease from $\sim 450 \arcsec$ (at \emph{GALEX} FUV 1516\,\AA) to $\sim 200 \arcsec$ (at 2MASS J $1.2\,\mu$m) and a subsequent increase to $\sim 300 \arcsec$ (at \emph{WISE} W2 $4.6\,\mu$m).
	
	The wavelength distribution for each of the \sersic{} index and the effective radius is well fitted with a quadratic function and a piecewise linear function with two slopes, respectively.\footnote{~The \emph{Spitzer}-IRAC 5.8 $\mu$m band is excluded from the polynomial fitting, because there is a large deviation in the decomposition at this band, which will be discussed in detail in Section \ref{Sec_Disc}.}  For the piecewise linear function, the turn point is set as a free parameter in the fit.  The extrema (or turn points) of these functions are generally in the wavelength range $\sim 1.0$--$1.5\,\mu$m.  The formulae of the best-fit curves are listed as follows:
	
	\begin{figure*}[!ht]
		\centering
		\plotone{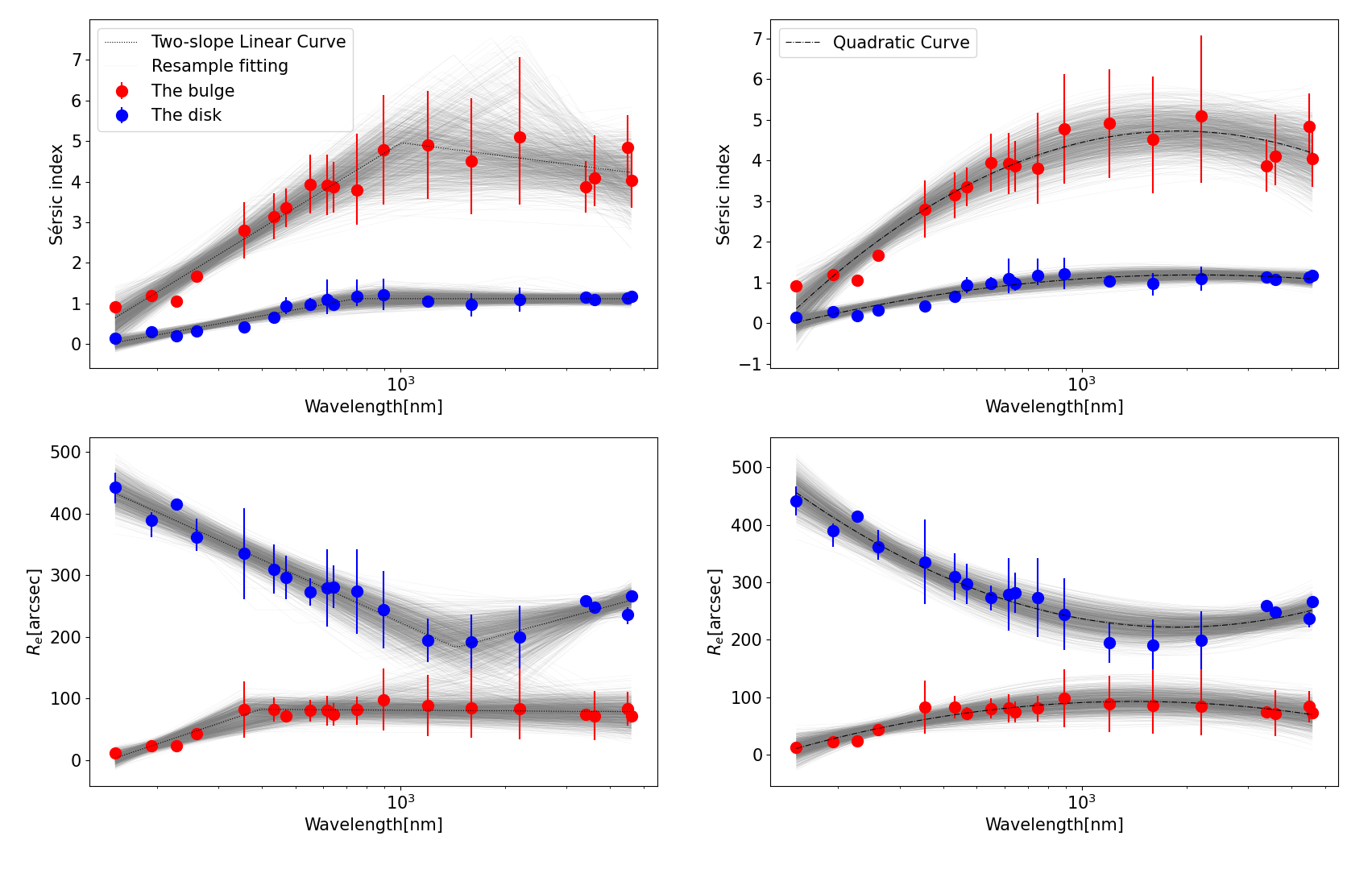}
		\caption{The \sersic{} index (top row) and the effective radius (bottom row) as a function of wavelength for the bulge (red filled circles) and the disk (blue filled circles), fitted with piecewise linear (left column) and quadratic (right column) curves.  In each panel, black dashed and dot-dashed lines represent the best-fit curves; gray solid lines represent fitting curves obtained from 1000 bootstrap resampling iterations, which define the error ranges for the fits.}
		\label{fig:final_err}
	\end{figure*}

        \onecolumngrid
	
	\begin{equation}\label{eq:n_bulge}
		n_\mathrm{~bulge} = (-3.63_{~\pm1.03})\log^2(\lambda)+(23.76_{~\pm6.07})\log(\lambda)+(-34.22_{~\pm8.66}),
	\end{equation}
	
	\begin{equation}\label{eq:n_disk}
		n_\mathrm{~disk} = (-0.88_{~\pm0.26})\log^2(\lambda)+(5.89_{~\pm1.53})\log(\lambda)+(-8.61_{~\pm2.18}),
	\end{equation}
	
	\begin{equation}\label{eq:Re_bulge}
		{R_e}_\mathrm{,~bulge} = (-86.94_{~\pm33.73})\log^2(\lambda)+(547.7_{~\pm202.22})\log(\lambda)+(-770.14_{~\pm291.55}),
	\end{equation}
	
	\begin{equation}\label{eq:Re_disk}
		{R_e}_\mathrm{,~disk} = (195.51_{~\pm48.13})\log^2(\lambda)+(-1280.48_{~\pm286.05})\log(\lambda)+(2318.48_{~\pm410.24}),
	\end{equation}
	
	\begin{equation}\label{eq:n_bulge_linear}
		n_\mathrm{~bulge} = \left\{
		\begin{aligned}
			& (5.25_{~\pm0.92})[\log(\lambda)-(3.00_{~\pm0.18})]+(4.96_{~\pm0.89}) , &&\quad \lambda \le (1000.00~_{\pm1.51})\,\mathrm{nm}\\
			& (-1.11_{~\pm2.64})[\log(\lambda)-(3.0_{~\pm0.18})]+(4.96_{~\pm0.89}) , &&\quad \lambda > (1000.00_{~\pm1.51})\,\mathrm{nm}\\
		\end{aligned}
		\right.
		, ~
	\end{equation}
	
	\begin{equation}\label{eq:n_disk_linear}
		n_\mathrm{~disk} = \left\{
		\begin{aligned}
			& (1.57_{~\pm0.25})[\log(\lambda)-(2.87_{~\pm0.10})]+(1.12_{~\pm0.16}) , &&\quad \lambda \le (741.31_{~\pm1.26})\,\mathrm{nm}\\
			& (-0.003_{~\pm 0.27})[\log(\lambda)-(2.87_{~\pm0.10})]+(1.12_{~\pm0.16}) , &&\quad \lambda > (741.31_{~\pm1.26})\,\mathrm{nm}\\
		\end{aligned}
		\right.
		, ~
	\end{equation}
	
	\begin{equation}\label{eq:Re_bulge_linear}
		{R_e}_\mathrm{,~bulge} =\left\{
		\begin{aligned}
			& (188.80_{~\pm 50.40})[\log(\lambda)-(2.60_{~\pm0.19})]+(82.38_{~\pm19.32}), &&\quad \lambda \le (398.11_{~\pm1.55})\,\mathrm{nm}\\
			& (-3.52_{~\pm 50.37})[\log(\lambda)-(2.60_{~\pm0.19})]+(82.38_{~\pm19.32}) , &&\quad \lambda > (398.11_{~\pm1.55})\,\mathrm{nm}\\
		\end{aligned}
		\right.
		, ~
	\end{equation}
	and
	\begin{equation}\label{eq:Re_disk_linear}
		{R_e}_\mathrm{,~disk} =\left\{
		\begin{aligned}
			& (-255.98_{~\pm43.7})[\log(\lambda)-(3.16_{~\pm0.12})]+(182.54_{~\pm28.59}) , &&\quad \lambda \le (1445.44_{~\pm1.32})\,\mathrm{nm}\\
			& (152.46_{~\pm99.77})[\log(\lambda)-(3.16_{~\pm0.12})]+(182.54_{~\pm28.59}) , &&\quad \lambda > (1445.44_{~\pm1.32})\,\mathrm{nm}\\
		\end{aligned}
		\right.
		, ~
	\end{equation}

        \twocolumngrid

	where $\lambda$ is the wavelength in units of nanometers (nm).  Uncertainties in the fitting functions are estimated by using the "bootstrap resampling" method (\citealt{Dey_2019}).  We artificially construct a sample containing 1000 mock galaxies randomly distributed within the error ranges for the M81 data at the 20 wavebands, fit the wavelength-dependent morphology of each mock galaxy in the same way, and hence obtain the best-fit curves for all of the mock galaxies.  The uncertainties in Equations \ref{eq:n_bulge}--\ref{eq:Re_disk_linear} are defined as standard deviation in the best-fit curves for the mock galaxies from that for M81.  The results of the "bootstrap resampling" method are displayed in Figure \ref{fig:final_err}.

	\subsection{Spectral Energy Distributions and Multiwavelength Flux Ratios}\label{Sec_Result_SED}
	
	Spectral energy distributions (SEDs) and multiwavelength flux ratios to the galaxy as a whole for the bulge and the disk are also displayed in Figure \ref{fig:final}.  Integrated measurements of the bulge, the disk, and the galaxy as a whole were performed by photometry with the same aperture.  The aperture is an ellipse centered at R.A. = 148$^\circ$.888, Dec. = 69$^\circ$.065, with the semi-major axis $750 \arcsec \times 450 \arcsec$ and the position angle $-20^\circ$.  Uncertainties in the measurements are estimated as a quadratic sum of calibration errors and background deviation.  The flux ratio is defined as the flux of the bulge or the disk (measured in the model image) divided by the flux of the galaxy as a whole (measured in the observed image) at a certain waveband.  The WIYN fluxes are not included in the SEDs and the flux ratios, because the field of view for the WIYN images is smaller than the photometric aperture.
	
	The observed SEDs for the bulge, the disk, and the galaxy as a whole were fitted with model spectra of stellar population synthesis by using the CIGALE software (\citealt{Boquien_2019}).  The initial guess of parameters is made in accordance with those in \citet{Nersesian_2019}.  In the SED fitting, we designate the solar metallicity ($Z$ = 0.02, \citealt{Kong_2000}) and the star formation history in an e-folding form on the star formation timescale 2600\,Myr which is the best-fit value for M81; stellar population age and interstellar dust attenuation are variable parameters.
	
	In the bottom-left panel of Figure \ref{fig:final}, the best-fit model spectra indicate that the mean age of stellar populations is 6 Gyr for the bulge, 2 Gyr for the disk, and 2.5 Gyr for the galaxy as a whole.  The color index NUV$-${\it r} is $\sim 4.02$ mag for the whole galaxy, $\sim 3.54$ mag for the disk, and $\sim 6.98$ mag for the bulge, respectively. Containing the oldest stellar population in the galaxy, the bulge is extremely red in color.
	
	\begin{figure*}[!ht]
		\centering
		\plotone{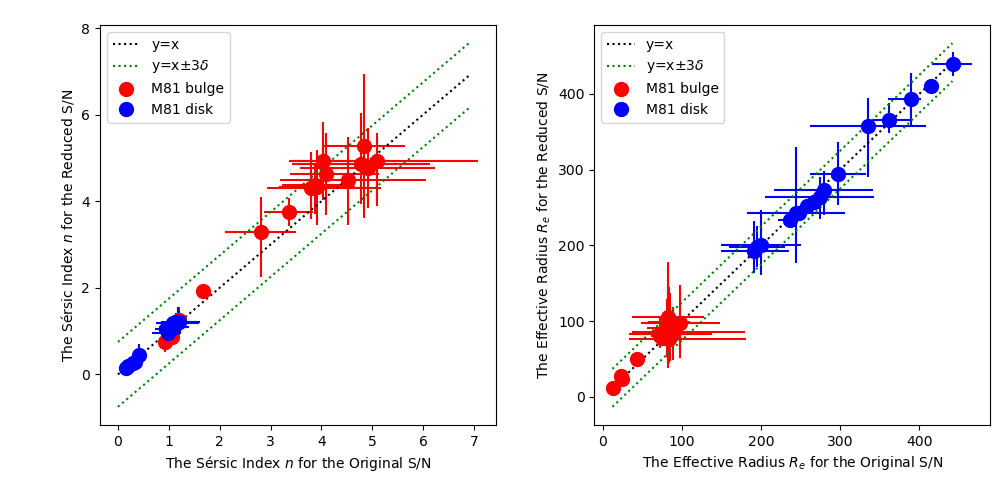}
		\caption{Comparisons between the same best-fit parameters in the images with the original and reduced S/N. By adding different levels of Gaussian noise with an mean value of zero, we unify all data into the same S/N as the \emph{GALEX} FUV band, which has the lowest S/N in all data. The left panel shows the comparison of the \sersic{} index $n$; the right panel shows the comparison of the effective radius $R_e$.  In each panel, symbols and lines are the same as assigned in Figure \ref{fig:goodness2sersic}.}
		\label{fig:lowsn}
	\end{figure*}
	
	The bottom-right panel of Figure \ref{fig:final} shows the bulge-to-galaxy and disk-to-galaxy flux ratios as a function of wavelength.  We can see that, at the UV bands, the disk dominates the galaxy ($>$ 80\% at Swift UVW2, GALEX NUV, and Swift UVW1), while the bulge is quite trivial ($<$ 10\% at GALEX FUV, Swift UVW2, and GALEX NUV); at longer wavelengths, the flux for the disk declines whereas that for the bulge rises, and the contributions of the bulge and the disk to the galaxy get close and comparable at optical and NIR bands ($\sim 50$\% at 3551--58000 \AA).

	\section{\textbf{DISCUSSION}}\label{Sec_Disc}
	
	In the above section, variations in a series of morphological parameters for the bulge and the disk in M81 with wavelength are displayed as a result of the structural decomposition by fitting the surface-brightness profiles.  In practice, goodness of the surface-brightness fit is related with several factors such as S/N of the data, the presence of the spiral arms, and the covariance between the parameters.  The potential influences on the fit, as well as implications of the results and comparisons with other studies, are presented and discussed in this section.
	
	\subsection{Sensitivity to the Signal-to-Noise Ratio}\label{Sec_Disc_SNR}
	
	Among the factors potentially affecting the surface-brightness fit, the noise level for the data often dominates the reliability of the products.  In this work, the multiwavelength images with different S/N are analyzed.  The difference in S/N is likely to contribute a systematic trend to the resultant wavelength-dependent morphology.  For the purpose of discerning the suspected imprint of S/N and eliminating its influence, we artificially impose noise on the images to equalize S/N at all of the wavebands, and then perform the same fit to compare the noisy results with the normal ones.
	
	\begin{figure*}[!ht]
		\centering
		\plotone{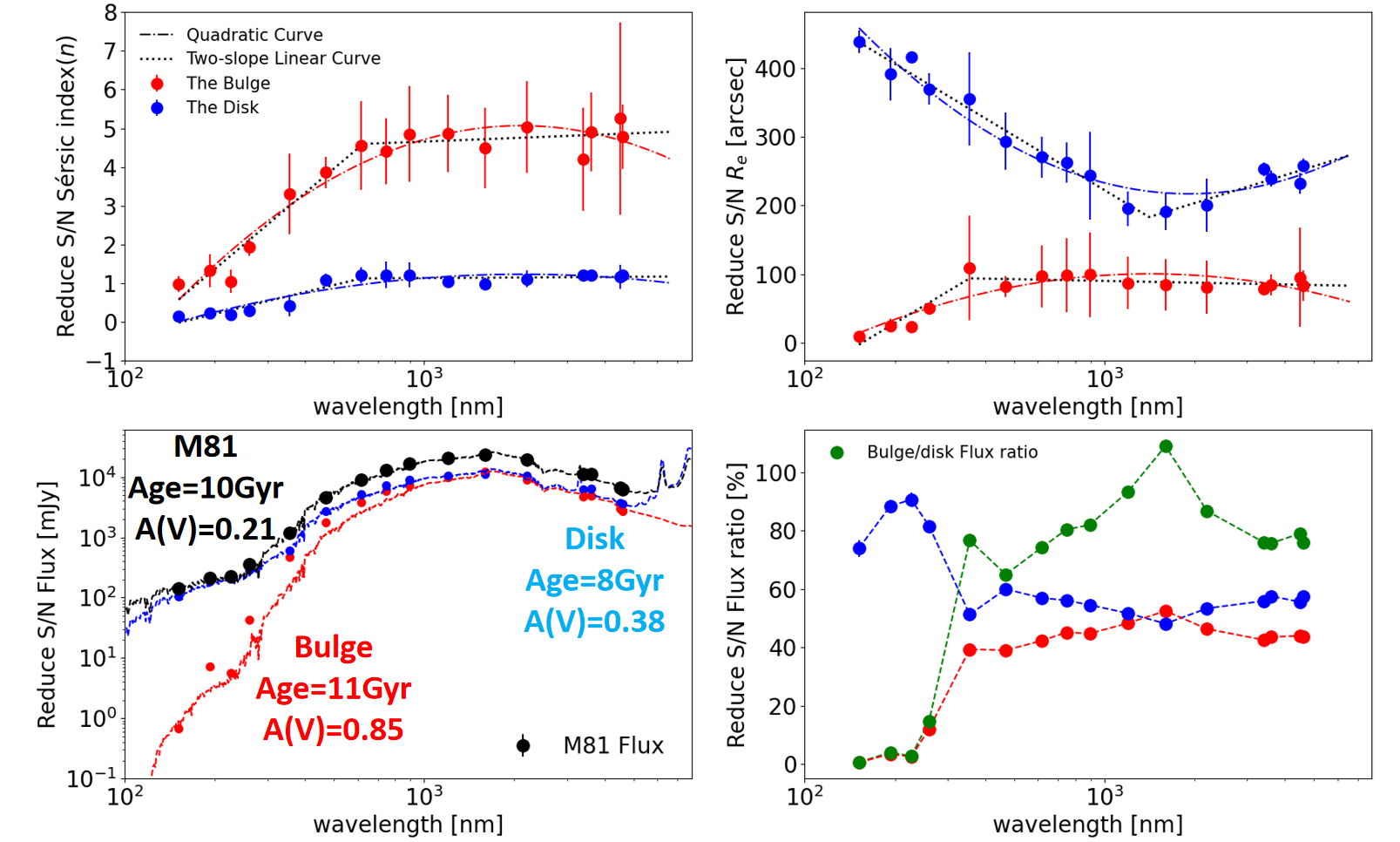}
		\caption{The same as Figure \ref{fig:final} but for all of the images with S/N reduced to an equal level.
		}
		\label{fig:snrvswave}
	\end{figure*}
	
	In accordance with the definition in \cite{Davari_2014}, S/N is calculated as follow:
	
	\begin{equation}\label{eq:S/N}
		\mathrm{\,S/N} = \frac{f}{\sqrt{f+A\sigma^2}} ~,
	\end{equation}
		
	where $f$ is the flux for the galaxy obtained with the aperture photometry, $A$ is the area of the aperture, and $\sigma$ is the background uncertainty.  The noise imposed on each image is configured to have a Gaussian distribution with a mean value of zero and different amounts of the standard deviation, so as to align S/N to that in the \emph{GALEX} FUV image ($\sim 90$, the lowest S/N among all of the data).\footnote{~In this step, we only add background noise and do not change the flux for the source.  The method for reducing S/N is addressed in \cite{Lotz_2008, Meert_2013, Davari_2014}.}  The WIYN and \emph{Spitzer}-IRAC 5.8 $\mu$m images are not included in this part of processing, because the WIYN fields of view are small, and the 5.8 $\mu$m band is contaminated by dust emission, which potentially results in an incorrect measurement of S/N.  As a consequence, there are 16 wavebands in total adopted for the investigation with the equalized S/N.  The imposition of the Gaussian noise inevitably introduce extra uncertainties in the results. In order to diminish the extra uncertainties, we carry out the alignment of S/N for 10 times and thereby create 10 images at each waveband; in each of the 10 images at each waveband, the surface-brightness profile is fitted to obtain the morphological parameters; the average of the 10 estimates is taken as the final value for each parameter at each waveband.  The parameters at the 16 wavebands with the equalized S/N are presented in Appendix \ref{Sec_Appendix_B}.
	
	\begin{figure*}[!ht]
		\centering
		\plotone{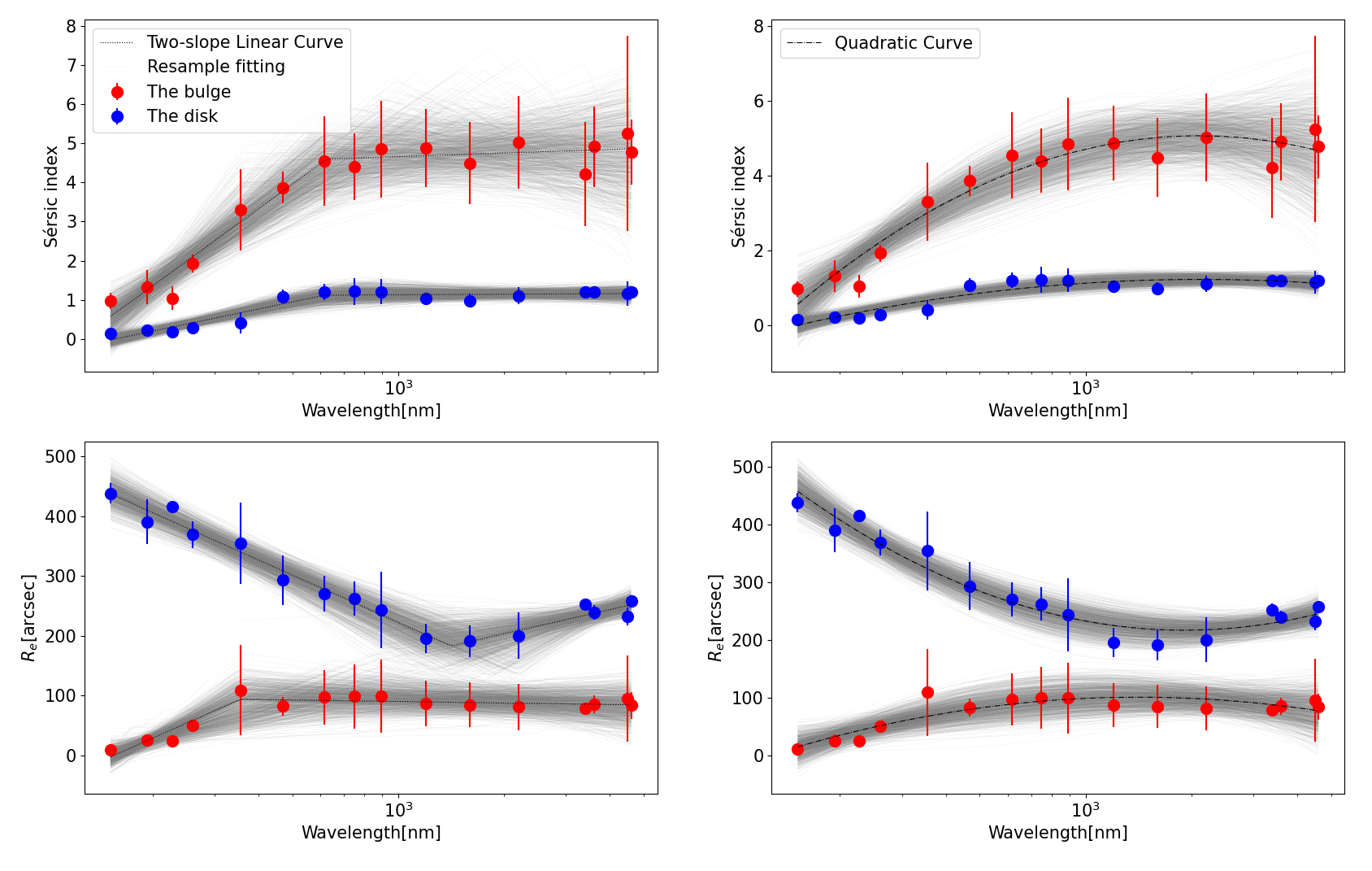}
		\caption{The same as Figure \ref{fig:final_err} but for the images with S/N reduced to an equal level.
		}
		\label{fig:snrvswave_err}
	\end{figure*}
	
	Figure \ref{fig:lowsn} shows the comparisons in the best-fit parameters between the original and processed S/N.  We can see that the S/N alignment does not affect most of the data points, but the \sersic{} index $n$ for the bulge at some NIR bands increases, albeit within the $3 \sigma$ range.  This behavior appears effective in the relation of the parameter with wavelength, as shown in Figures \ref{fig:snrvswave} and \ref{fig:snrvswave_err}.  We still adopt a quadratic curve and a two-slope piecewise linear curve to, respectively, fit each of the parameter-wavelength relations.  The formulae with uncertainties in coefficients are listed as follows:
	
    \onecolumngrid
	\begin{equation}\label{eq:snr_n_bulge}
		n_\mathrm{~bulge} = (-3.45_{~\pm1.50})\log^2(\lambda)+(22.94_{~\pm8.48})\log(\lambda)+(-33.05_{~\pm11.67}),
	\end{equation}
	
	\begin{equation}\label{eq:snr_n_disk}
		n_\mathrm{~disk} = (-0.93_{~\pm0.31})\log^2(\lambda)+(6.19_{~\pm1.81})\log(\lambda)+(-9.08_{~\pm2.55}),
	\end{equation}
	
	\begin{equation}\label{eq:snr_Re_bulge}
		{R_e}_\mathrm{,~bulge} = (-91.26_{~\pm55.16})\log^2(\lambda)+(575.35_{~\pm317.53})\log(\lambda)+(-806.12_{~\pm440.99}),
	\end{equation}
	
	\begin{equation}\label{eq:snr_Re_disk}
		{R_e}_\mathrm{,~disk} = (198.33_{~\pm46.68})\log^2(\lambda)+(-1302.23_{~\pm276.36})\log(\lambda)+(2354.71_{~\pm394.79}),
	\end{equation}
	
	\begin{equation}\label{eq:snr_n_bulge_linear}
		n_\mathrm{~bulge} = \left\{
		\begin{aligned}
			& (6.53_{~\pm1.47})[\log(\lambda)-(2.79_{~\pm0.17})]+(4.59_{~\pm0.81}) , &&\quad \lambda \le (616.59_{~\pm1.48})\,\mathrm{nm}\\
			& (0.30_{~\pm2.66})[\log(\lambda)-(2.79_{~\pm0.17})]+(4.59_{~\pm0.81}) , &&\quad \lambda > (616.59_{~\pm1.48})\,\mathrm{nm}\\
		\end{aligned}
		\right.
		, ~
	\end{equation}
	
	\begin{equation}\label{eq:snr_n_disk_linear}
		n_\mathrm{~disk} = \left\{
		\begin{aligned}
			& (1.87_{~\pm0.31})[\log(\lambda)-(2.79_{~\pm0.09})]+(1.12_{~\pm0.17}) , &&\quad \lambda \le (616.59_{~\pm1.23})\,\mathrm{nm}\\
			& (0.05_{~\pm0.28})[\log(\lambda)-(2.79_{~\pm0.09})]+(1.12_{~\pm0.17}) , &&\quad \lambda > (616.59_{~\pm1.23})\,\mathrm{nm}\\
		\end{aligned}
		\right.
		, ~
	\end{equation}
	
	\begin{equation}\label{eq:snr_Re_bulge_linear}
		{R_e}_\mathrm{,~bulge} =\left\{
		\begin{aligned}
			& (260.81_{~\pm88.19})[\log(\lambda)-(2.55_{~\pm0.19})]+(94.15_{~\pm25.83}) , &&\quad \lambda \le (354.81_{~\pm1.55})\,\mathrm{nm}\\
			& (-21.70_{~\pm44.26})[\log(\lambda)-(2.55_{~\pm0.19})]+(94.15_{~\pm25.83}) , &&\quad \lambda > (354.81_{~\pm1.55})\,\mathrm{nm}\\
		\end{aligned}
		\right.
		, ~
	\end{equation}
	and
	\begin{equation}\label{eq:snr_Re_disk_linear}
		{R_e}_\mathrm{,~disk} =\left\{
		\begin{aligned}
			& (-262.32_{~\pm41.73})[\log(\lambda)-(3.14_{~\pm0.10})]+(183.17_{~\pm22.19}) , &&\quad \lambda \le( 1442.54_{~\pm1.26})\,\mathrm{nm}\\
			& (134.34_{~\pm77.55})[\log(\lambda)-(3.14_{~\pm0.10})]+(183.17_{~\pm22.19}) , &&\quad \lambda > (1442.54_{~\pm1.26})\,\mathrm{nm}\\
		\end{aligned}
		\right.
		, ~
	\end{equation}
	
        \twocolumngrid

	where $\lambda$ is the wavelength in units of nanometers (nm).  After reducing S/N to an equal value for all of the images, the \sersic{} index as a function of wavelength for the bulge appears with no peak and is better fitted by a two-slope piecewise linear curve.  At the wavelength $\lambda \leq 6252$\,\AA~according to the two-slope piecewise linear curve (or shorter than SDSS {\it r} band according to the data), it increases sharply with wavelength, while over the turn-point wavelength, the trend becomes nearly flat.  In contrast to the \sersic{} index for the bulge, the other parameters stay the same as before.
	
	It should be noted that, the S/N equalization is conducted for the purpose of inspecting the sensitivity to S/N; the result of this processing does not mean that it is closer to the real nature than that of keeping S/N unchanged.  An increment of noise will cause a random change in brightness for galactic pixels, which hence changes the \sersic{} index for the bulge at the sensitive wavebands by random amounts.  That is the reason why we make 10 sets of parameters and take the averages.  However, there remains to be fluctuation in the wavelength-dependent morphology due to the random effect, which can be hardly eliminated outright.  Notwithstanding, the result suffices to illustrate the sensitivity of the wavelength-dependent morphology to S/N.  According to our analysis, the \sersic{} index with a larger value, i.e., a surface-brightness profile with higher concentration, is more delicate and susceptible to background noise.  As a consequence, most of the parameters are robust against the change in background noise, but the \sersic{} index for the bulge at the NIR bands are affected.  A larger amount of noise imposition tends to cause a higher value for the \sersic{} index.  Nevertheless, the influence seems not too disruptive; the difference is less than the uncertainty in the estimate, and the change in the relation of the \sersic{} index with wavelength for the bulge is modest.  In the wavelength range from optical to NIR, it would still be arbitrary to claim a constant value for the \sersic{} index, but it is convincing that the variation in the \sersic{} index (if any) is much shallower compared with the sharp rise from FUV to optical bands.  A more solid and general conclusion will rest on studies of a large sample of galaxies.
		
	\begin{figure*}[!ht]
		\centering
		\plotone{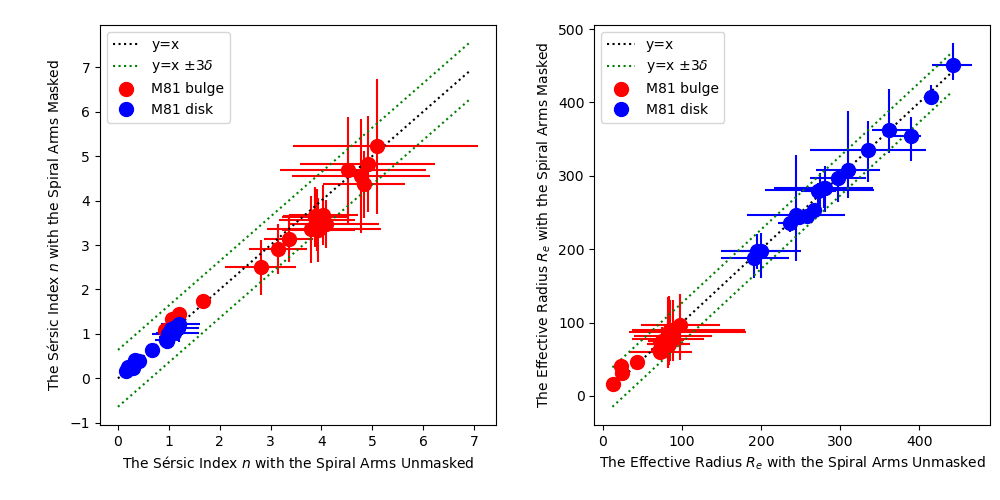}
		\caption{Comparisons between the same best-fit parameters with the spiral arms masked and unmasked.  The left panel shows the comparison of the \sersic{} index $n$; the right panel shows the comparison of the effective radius $R_e$.  In each panel, red solid circles represent the bulge, and blue solid circles represent the disk; black dotted line represents the line of unity ($y = x$), with green dotted lines enveloping the range of three times the standard deviation ($y = x \pm 3 \delta$, where $\delta$ is the standard deviation in the data).  Error bars are superimposed on the data points.
		}
		\label{fig:goodness2sersic}
	\end{figure*}
	
	\subsection{Impact of the Spiral Arms}\label{Sec_Disc_SpA}
	
	In this work, we use two \sersic{} components to reproduce the bulge and the disk in M81, with the spiral arms left unfitted.  Compared with the bulge and the disk, the spiral arms are an asymmetric structure and hence not able to be reproduced by a common model.  In most of the residual images (see Figures \ref{fig:fd}--\ref{fig:I3} in Appendix \ref{Sec_Appendix_A}), particularly at short wavelengths, spiral shapes are clearly seen.  In this case, it is necessary to understand the impact of the spiral arms on the morphological parameterization of the bulge and the disk.  With this in mind, we masked the spiral arms and then performed the fit for comparison with the results of the spiral arms unmasked.\footnote{~The spiral arms are defined in the residual images as the pixels with the brightness larger than the 5$\sigma$ level.}
	
	Figure \ref{fig:goodness2sersic} shows the comparisons between the same best-fit parameters for the spiral arms masked and unmasked.  The differences, as seen from this figure, are less than three times the standard deviation at all wavelengths, which illustrates that the presence of the spiral arms has a very trivial impact on the bulge-disk decomposition, and the two-\sersic{} fit suffices to characterize the morphology of the bulge and the disk in M81.
	
	Independent evidence proving that modeling spiral arms are not crucial has been provided by \cite{Gao_2017} through detailed morphological analysis of six spiral galaxies.  They have found that the presence of spiral arms has a subtle impact on the parameterization of bulges; specifically in the \sersic{} index, it introduces an uncertainty of $\sim 10$\% for unbarred galaxies and $\sim 1$\% for barred galaxies.  In addition, \cite{Gong_2023} have elaborately investigated the \emph{Spitzer}-IRAC 4.5 $\mu$\,m image for M81 by adopting six types of decompositions with the number of model components from 1 to 5.  The most complicated model contains a bulge (fitted with one \sersic{} function), a disk (fitted with one \sersic{} function), a pair of outer spiral arms (fitted with one coordinate-rotation truncated \sersic{} function), a pair of inner spiral arms (fitted with one coordinate-rotation truncated \sersic{} function), and a galactic nucleus (fitted with one PSF).  This experiment  that, as long as both of the bulge and the disk are included in the model (i.e., with the two-\sersic{} functions),\footnote{For the galaxy M81, as well as any galaxy hosting an AGN, a PSF for fitting the nucleus is also requisite.} the results are relatively stable.
	
	Indeed, it would be ideal to take spiral arms as well as other substructures (if any) into model components, but this would induce extra errors and consume more time in the fit.  In general cases, two \sersic{} components are an optimal mode for describing morphology of galaxies even if asymmetric substructures such as spiral arms, bars, or rings exist, especially when studying a large sample of galaxies.
	
	\subsection{Degeneracy in the Fit}\label{Sec_Disc_nRecoVar}
	
	\begin{figure*}[!ht]
		\centering
		\plotone{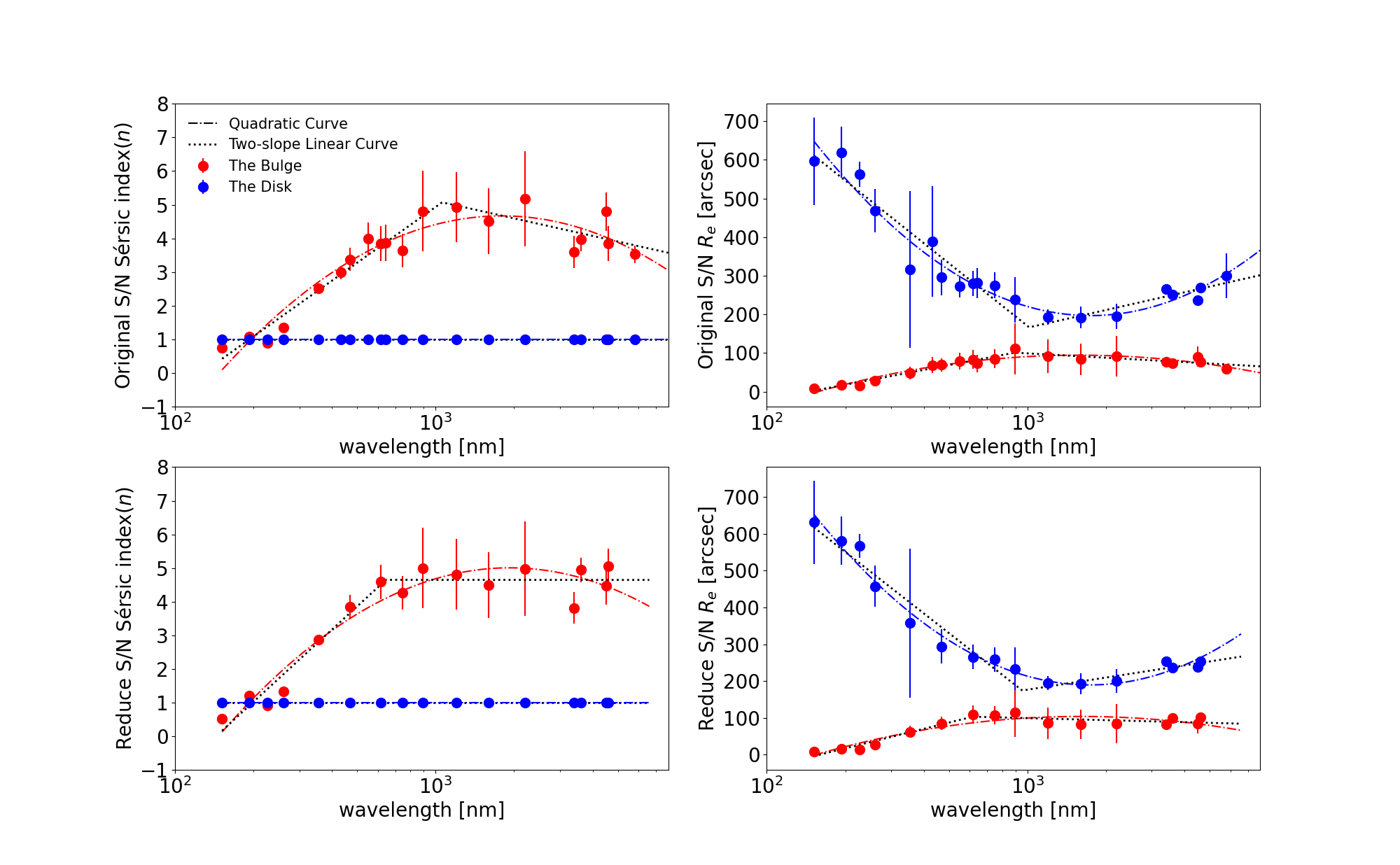}
		\caption{The \sersic{} index (left column) and the effective radius (right column) as a function of wavelength for the bulge (red filled circles) and the disk (blue filled circles) with original (top row) and reduced S/N (bottom row).  Dot-dashed and dotted lines represent the best-fit quadratic and piecewise linear curves, respectively.  The \sersic{} index for the disk is fixed into $n_\mathrm{~disk} = 1$ (the exponential disk).}
		\label{fig:disnfixed}
	\end{figure*}
	
	Degeneracy of free parameters seems always to be an intractable problem in any kind of fits.  In this work, we adopt two \sersic{} components to fit the bulge and the disk in M81, respectively, where the \sersic{} index and the effective radius are free parameters when fitting surface-brightness profiles.  This part of the discussion focuses on the degeneracy in the surface-brightness fit.
	
	\begin{figure*}[h]
		\centering
		\plotone{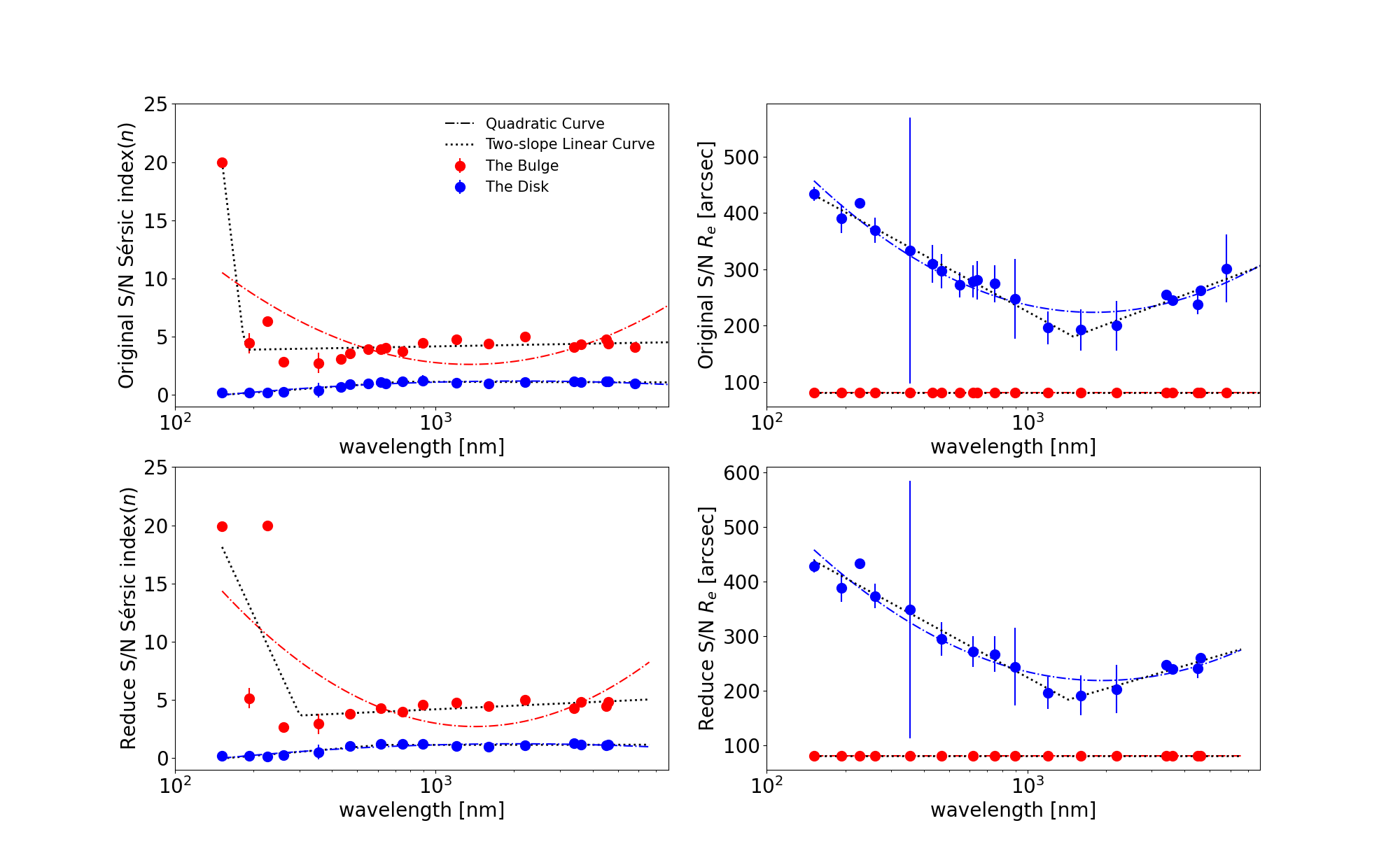}
		\caption{The same as Figure \ref{fig:disnfixed} but with the effective radius for the bulge fixed to ${R_e}_\mathrm{,~bulge} = 80\arcsec.71$ (the mean value at all of the wavebands apart from the four UV bands).}
		\label{fig:bulgerefixed}
	\end{figure*}
	
	\begin{figure*}[h]
		\centering
		\plotone{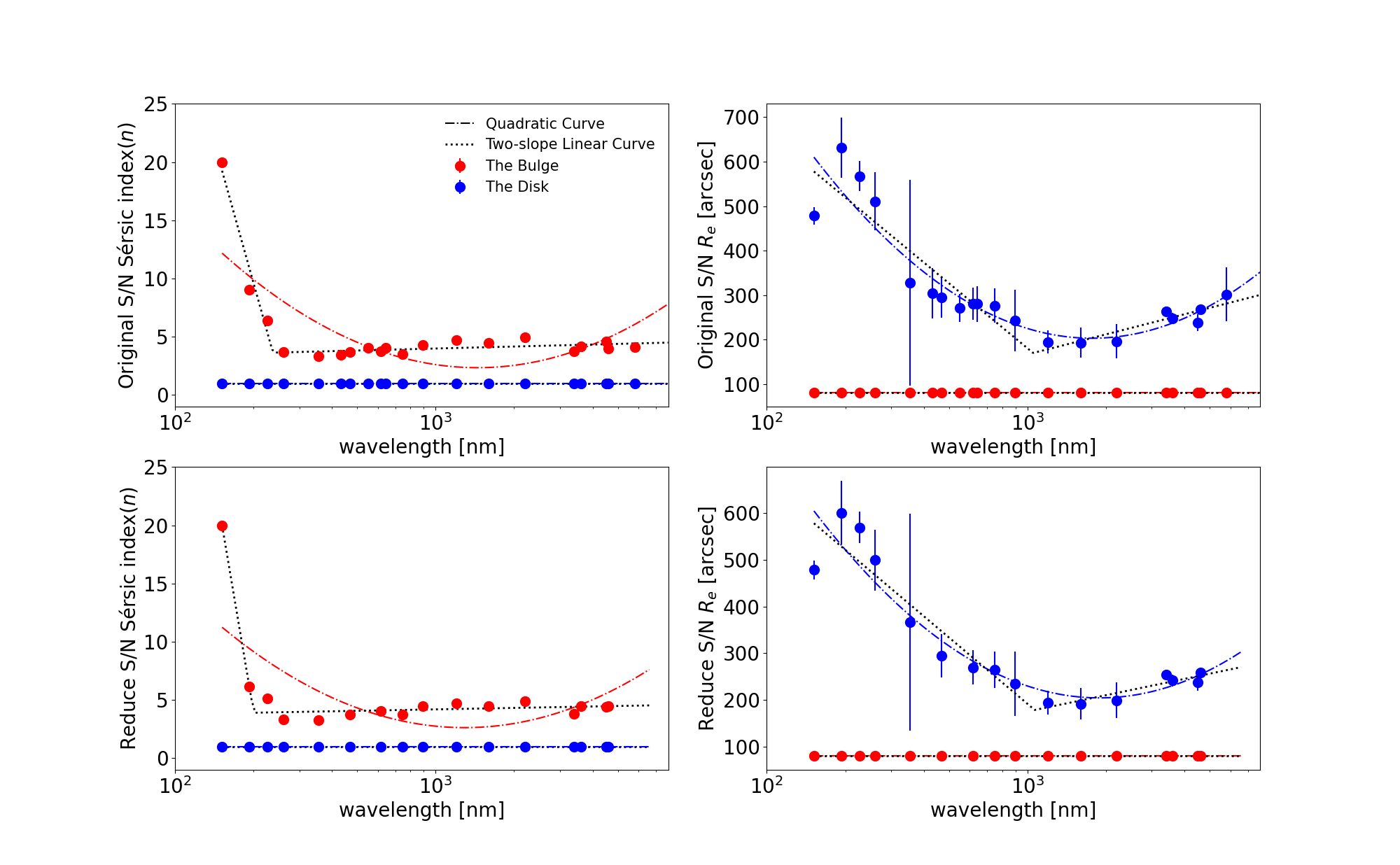}
		\caption{The same as Figure \ref{fig:disnfixed} but with the \sersic{} index for the disk fixed to $n_\mathrm{~disk} = 1$ (the exponential disk) and the effective radius for the bulge fixed to ${R_e}_\mathrm{,~bulge} = 80\arcsec.71$ (the mean value at all of the wavebands apart from the four UV bands) at the same time.}
		\label{fig:disknandbulgerefixed}
	\end{figure*}
	
	With the \sersic{} function adopted in the fit, the bulge and the disk share a common space from the center to infinity, and thus the two components always overlap with each other.  From Figures \ref{fig:final}, \ref{fig:final_err}, \ref{fig:snrvswave}, and \ref{fig:snrvswave_err}, we can see that the \sersic{} index for the disk and the effective radius for the bulge appears constant with wavelength except for the UV bands, where the parameters appear to have lower values, which hints at possible degeneracy of the two components at these bands.  In order to eliminate the possibility, we fix the \sersic{} index for the disk into $n_\mathrm{~disk} = 1$ (the exponential disk), the effective radius for the bulge into the mean value of all of the wavebands apart from UV (${R_e}_\mathrm{,~bulge} = 80\arcsec.71$), and both at the same time.  The results are shown in Figures \ref{fig:disnfixed}--\ref{fig:disknandbulgerefixed}, respectively.  These figures illustrate that fixing the parameters has almost no effect on the fit but cause a problematic estimate at the UV bands, which demonstrates that the bulge and the disk are not heavily coupled but separable by the decomposition.
	
	\begin{figure*}[ht]
		\centering
		\plotone{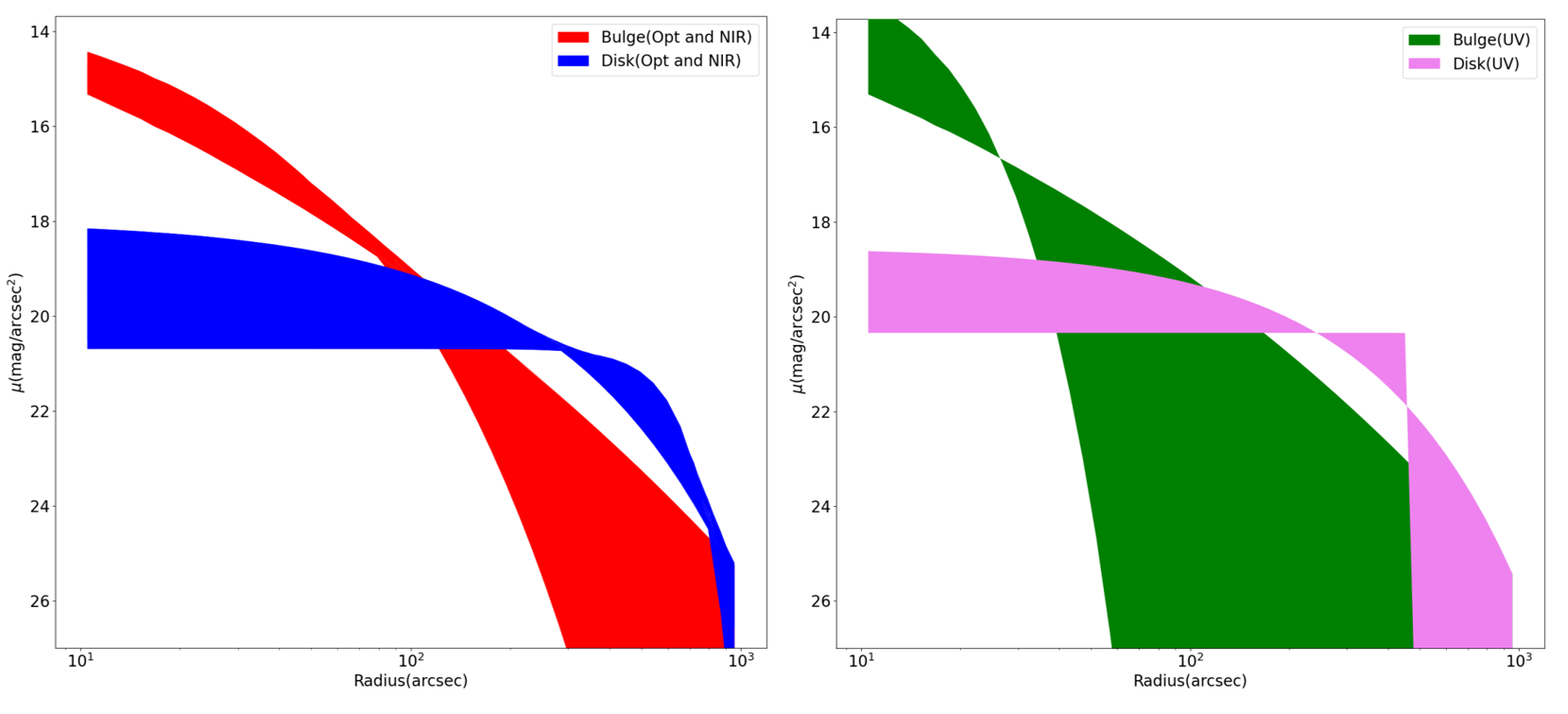}
		\caption{Variations in surface-brightness profiles for the artificially generated bulge and disk in the wavelength range from \emph{GALEX} FUV to \emph{WISE} W2, by using Equations \ref{eq:n_bulge}--\ref{eq:Re_disk}.   The colored zones represent the coverage of these multiwavelength profiles, and indicate the effect of the changes in the \sersic{} index and the effective radius on the profiles.  The profiles are reproduced on the assumption of a constant total magnitude for both of the bulge and disk.}
		\label{fig:conva}
	\end{figure*}
	
	For an individual component, since the \sersic{} index and the effective radius are free parameters, the two parameters are necessarily coupled, which appears as comparative covariance between the two parameters, but the surface-brightness profile generally keeps unchanged in shape.  In order to inspect the degeneracy, a series of models for the bulge and the disk with the same total brightness are artificially generated by use of Equations \ref{eq:n_bulge}--\ref{eq:Re_disk} in the wavelength range for our data.  By following the approach in \cite{Kelvin_2012}, in Figure \ref{fig:conva} we show variations in surface-brightness profiles as the \sersic{} index and the effective radius jointly vary.  In this figure, the optical--NIR bands and UV bands are separately displayed to show the distinct behaviors between the two wavelength parts.  The ranges for the \sersic{} index and the effective radius are regulated in accordance with the results in Figure \ref{fig:final}.  At the optical--NIR bands, the bulge model ranges 2.60--5.37 in the \sersic{} index and 49$\arcsec$.71--91$\arcsec$.87 in the effective radius; the disk model ranges 0.51--1.19 in the \sersic{} index and 222$\arcsec$.90--353$\arcsec$.31 in the effective radius.  At the UV bands, the bulge model ranges 0.65--2.26 in the \sersic{} index and 10$\arcsec$.54--42$\arcsec$.89 in the effective radius; the disk model ranges 0.014--0.42 in the \sersic{} index and 371$\arcsec$.06--453$\arcsec$.38 in the effective radius.
	
	It is obvious from Figure \ref{fig:conva} that the bulge and the disk differ quite a lot in surface-brightness profiles, which indicates a good separation between the two components in the decomposition by the fit.  For individual \sersic{} components, the surface-brightness profiles for the bulge at the optical--NIR bands and the disk at the UV bands are relatively stable with a mild variation, as the \sersic{} index and the effective radius vary in a wide range, whereas the UV bulge covers a vast area in the surface-brightness plane indicating an enormous change in the profiles.  These behaviors imply that covariance between the \sersic{} index and the effective radius exists in the optical--NIR bulge and the UV disk, where, comparatively, changes in the two parameters are likely to have no effect on the surface-brightness profiles.  This prevents accurate measurements of the galactic morphology by means of the surface-brightness fits.  By contrast, the UV bulge appears with the surface-brightness profile depending more on the morphological parameters, and the surface-brightness fits are an effective tool for determining the \sersic{} index and the effective radius.

	\subsection{Morphology of the Bulge}\label{Sec_Disc_Bul}
	
	\subsubsection{Multiwavelength Features}\label{Sec_Disc_Bul_MultiW}
	
	It is interesting for the bulge of M81 that the \sersic{} index ($n$) varies widely with wavelength, steeply from $n \sim 1$ to $n \sim 4$ in the UV--optical range and shallowly between $n \sim 4$--$5$ at the optical--NIR bands.  In general cases, the \sersic{} index is an identification of classical bulges or pseudo-bulges.  The \sersic{} index equal to 2 is the usual line of demarcation between classical bulges ($n > 2$) and pseudo-bulges ($n \leq 2$, \citealt{Fisher_2008}).\footnote{~In addition to the \sersic{} index, there are also some alternative schemes, such as the Petrosian concentration index, the Kormendy relation, and kinematic indices (i.e., rotation velocity and velocity dispersion), for distinguishing between classical and pseudo-bulges (\citealt{Gadotti_2009, Neumann_2017, Gao_2020, Gao_2022}).  We focus the \sersic{} index in this work.  Readers are suggested to take multiple criteria for precise classification of bulges.}  The wavelength-dependent $n$ manifests that the bulge of M81 appears as a prominently classical bulge at the optical and NIR bands but as a bulgeless system at the UV bands.  This hints that if we observe through UV channels, we are likely to mistake a classical bulge for a pseudo-bulge or no bulge.  In this case, it is possible for one galaxy to be classified into different morphological types if observed at different wavebands, particularly for early-type spiral galaxies of which the UV morphology usually behaves like late types (\citealt{Kuchinski_2000}).
	
	The variation in the morphology of bulges with wavelength provides an interpretation of the inconsistency between observations of high-redshift galaxies and modeling of galaxy evolution. It has been reported that bulges are predicted to have been formed at the redshift $z \sim 2$, but galaxies at $z$ = 1--3 are found to be mainly dominated by disks (\citealt{Dokkum_2013, Bruce_2014, Shibuya_2015, Toft_2017, Dimauro_2019, Tacchella_2019, Hashemizadeh_2022arXiv}).  In practical terms, for high-redshift galaxies, optical telescopes observe their rest-frame UV appearances.  However, according to the increasing trend of the \sersic{} index with wavelength discovered in this work, a classical bulge becomes identifiable (i.e., $n \sim 3$) at the wavelength $\lambda \sim 3551$\,\AA~(at the SDSS u band), prominent (i.e., $n \sim 4$) at $\lambda \sim 5500$\,\AA~(at the WIYN V band), and possesses the \sersic{} index in a constant range (i.e., $n \sim 4$--$5$) from optical V to NIR J, H, K$_s$, and longer bands.  This is to say, if a galaxy with the inherent morphology similar to M81 is located at the redshift $z = 2$, then it will be identified with the \sersic{} index $n > 3$ at the observed wavelength $\lambda > 1.0\,\mu$m (corresponding to the rest-frame wavelength $\lambda > 3551$\,\AA) and $n > 4$ at the observed wavelength $\lambda > 1.6\,\mu$m (corresponding to the rest-frame wavelength $\lambda > 5500$\,\AA).  In this case, morphological classifications based on optical observations through $\lambda \leq 1.0\,\mu$m can hardly identify the presence of classical bulges but likely mistake early-type spiral galaxies for late-type ones at redshifts $z \geq 2$.  For the redshift $z \geq 3$, the identification of the classical bulge by using the \sersic{} index rests on observations beyond NIR J, H, K$_s$ bands, for which ground-based and last-generation space telescopes fail to achieve, whereas the James Webb Space Telescope (\emph{JWST}) meets the requirement and has great potential to make significant discoveries on this topic (\citealt{Ferreira_2022}).
	
	The SED for the bulge (in the bottom-left panel of Figure \ref{fig:final}) indicates that it is an old system ($\sim 6$\,Gyr).  The aged stellar population radiates dominantly at optical and NIR bands, but dimly at UV bands.  Taking 100\,Myr to divide stellar populations into old ($\geq 100$\,Myr) and young ($< 100$\,Myr) groups by following \cite{Verstocken_2020}, the SED-fitting result shows that the mass ratio of young stars in the bulge to the total stellar mass is $4 \times 10^{-5}$, while the luminosity ratio of young stars in the bulge to the total stellar luminosity is $5 \times 10^{-3}$.  For the disk, the mass ratio of young stars to the total stellar mass is $3 \times 10^{-3}$, while the luminosity ratio of young stars to the total stellar luminosity is 0.23.  This means that there are almost no young stars in the bulge, while the disk inhabits both young and old stars.
	
	\begin{figure*}[!ht]
		\centering
		\plotone{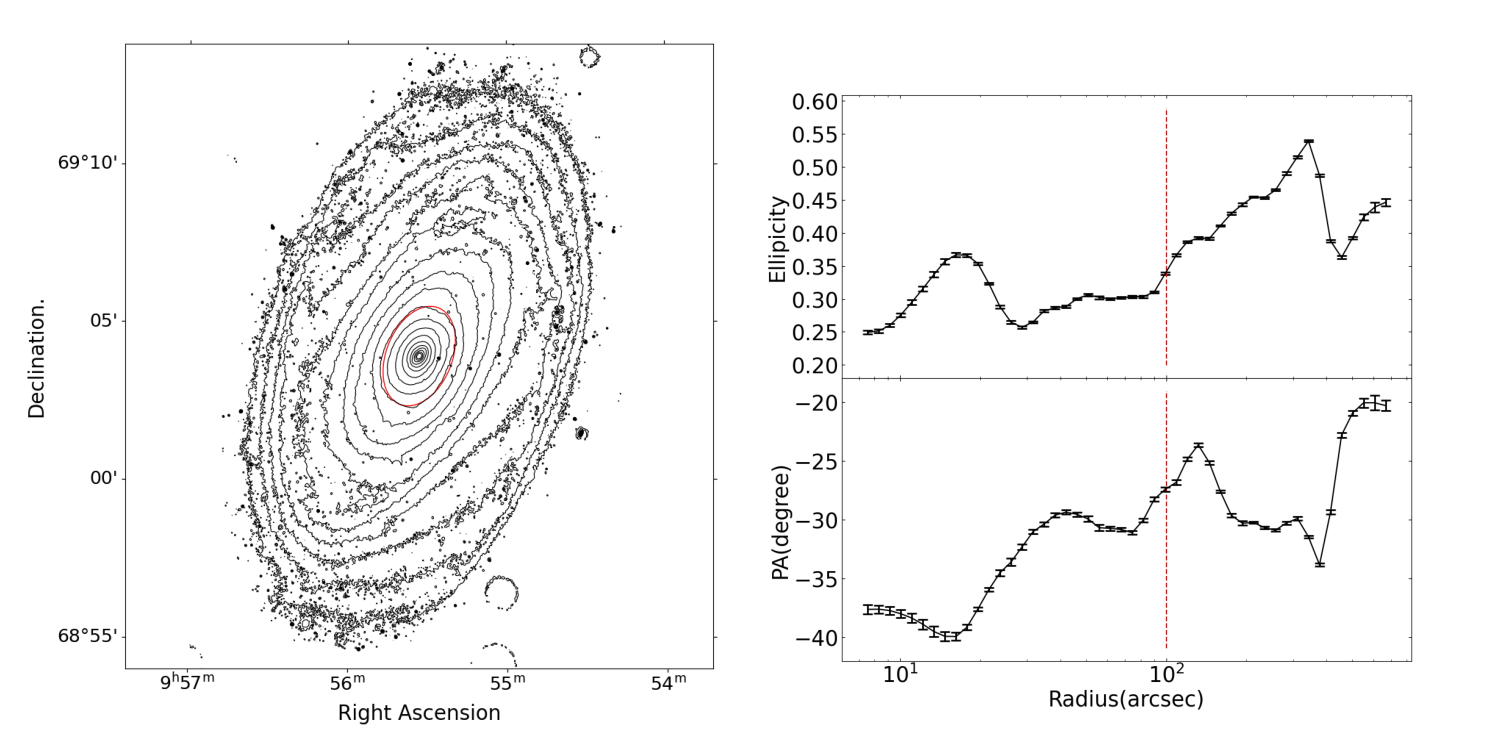}
		\caption{Left panel: isophote for M81 in the SDSS-{\it i} band image on a logarithmic scale, where the red ellipse marks a radial distance of 100$\arcsec$.  Right panels: radial profiles of ellipticity (upper panel) and position angle (lower panel) in the isophote.  The red dotted line indicates a radius of 100$\arcsec$.}
		\label{fig:sdssgiso}
	\end{figure*}
	
	\subsubsection{Lack of Pseudo-bulge Features}\label{Sec_Disc_Bul_Psd}
	
	It is worth noting that, classical bulges and pseudo-bulges are likely to coexist in galaxies (\citealt{Erwin_2003, Athanassoula_2005, Fisher_2010, Luo_2020, Yu_2022}).  If a pseudo-bulge coexists with the classical bulge in M81, then the variation in the morphology with wavelength for the bulge is interpretable with the coexistence of a classical bulge (with the \sersic{} index $n \geq 3$, containing old stellar populations, dominating optical and NIR bands) and a pseudo-bulge (with the \sersic{} index $n \sim 1$, containing younger stellar populations, dominating UV bands).  In this case, the low value for the \sersic{} index at the UV bands would be an impact of the pseudo-bulge.  In this section, we examine the possibility of the presence of a pseudo-bulge in M81.
	
	The presence of a pseudo-bulge, overlapping with a classical bulge, is likely to leave a signature in isophote for the galaxy (\citealt{Erwin_2015}).  By inspecting the ellipticity and the position angle of the isophote, we are able to suggest whether or not a pseudo-bulge coexists with the classical bulge.  In this work, the isophote is created in the SDSS-{\it i} band image, where the maximum is the highest flux in the image, and the minimum is the $3\sigma$ background deviation.  There are 20 contours in total, on a logarithmic scale, in the isophote.  Figure \ref{fig:sdssgiso} shows the result of the isophote and relevant analysis.
	
	During the isophote inspection, we define the bulge-dominated area at radii $\leq 100 \arcsec$, and the disk-dominated area at radii $> 100 \arcsec$, since at a radius of $\sim 100 \arcsec$, the bulge and the disk are equal in surface brightness in the SDSS-{\it i} image (as can be seen from Figure \ref{fig:i} in Appendix \ref{Sec_Appendix_A}).  The isophote map in Figure \ref{fig:sdssgiso} exhibits that the shape of the contours for the bulge-dominated area appears much rounder than that for the disk-dominated area.  The ellipticity of the isophote is in the range 0.18--0.38 for the bulge-dominated area and 0.35--0.58 for the disk-dominated area.\footnote{~The hump feature in the ellipticity profile around a radius of $\sim 16\arcsec$ in Figure \ref{fig:sdssgiso} is caused by small-scale structures inside the bulge (\citealt{Elmegreen_1995}; Mao et al., in preparation).}  The ellipticities for both of the bulge- and disk-dominated areas are consistent with those for the bulge and disk models in the morphological decomposition.  In the ellipticity profiles, the bulge-dominated area is systematically lower than the disk-dominated area.  The low ellipticity for the bulge-dominated area manifests absent of the pseudo-bulge.  Otherwise, if there is a pseudo-bulge, the bulge-dominated area (at least for its outer part) in the isophote will have high ellipticity, even comparable with the disk (\citealt{Erwin_2015}).
	
	In addition to the isophote analysis, a number of alternative diagnostics, such as the Petrosian concentration index (\citealt{Gadotti_2009, Neumann_2017}), the Kormendy relation (\citealt{Gao_2020}), and kinematic indices (i.e., rotation velocity and velocity dispersion, \citealt{Kormendy_1982}), indicate the presence of a classical bulge and no sign of a pseudo-bulge for M81 from other points of view.\footnote{~On the basis of preliminary analysis of the imaging data at SDSS-{\it i} band, we obtained the Petrosian concentration index and the Kormendy relation for the bulge of M81, which shows that it lies in a typical range for classical bulges; on the other hand, according to the spectroscopic data in \cite{Kormendy_1982}, the bulge in M81 appears with a rotation velocity less than velocity dispersion, which is a typical behavior of a classical bulge with no pseudo-bulge feature from a kinematic point of view.}
	
	The lack of pseudo-bulge features makes the nature of the exponential-like distribution for the UV bulge (i.e., to be more precise, the compact exponential UV component) questionable.  The most likely reason is a contribution of the disk to the central area.  The bulge and the disk overlap with each other.  The relative intensity of the bulge to the disk varies with wavelength, and at UV bands the disk dominates the galaxy in flux and morphology.  In this case, the prominent disk is likely to contribute to the bulge component during the decomposition.  From the SEDs for the bulge in Figures \ref{fig:final} and \ref{fig:lowsn}, we can see a systematic offset of the data points toward higher fluxes from the best-fit SEDs at the UV--optical bands.  This illustrates a flux excess for the bulge component, which is a probable aftermath of the disk contribution.  Nevertheless, we still cannot rule out the other possibility that the compact exponential UV component is exactly the classical bulge imaged through UV channels, even while accounting for the disk contribution.  Many studies have disclosed some old elliptical/spheroidal systems with a considerable amount of UV emission (\citealt{OConnell_1992, Kuchinski_2000, Marcum_2001, Windhorst_2002}).  Despite the reason for the exponential UV component, it is most important that we are quite likely to misidentify the morphological type of galaxies (i.e., to mistake a classical bulge for a pseudo-bulge or a no-bulge system by observing through UV channels), which often takes place in high-redshift studies.
	
	\begin{figure*}[!ht]
		\centering
		\plotone{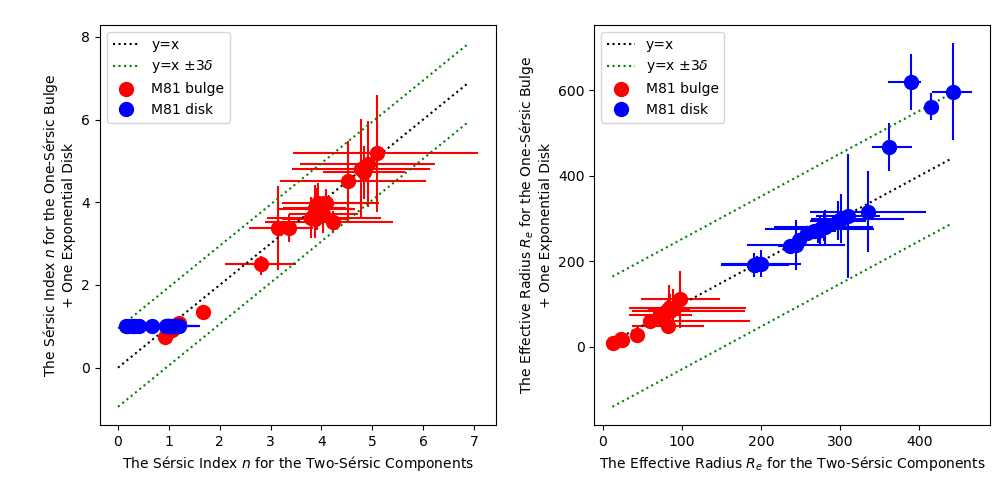}
		\caption{Comparisons between the same best-fit parameters for the two \sersic{} components and the one-\sersic{} bulge plus one-exponential (i.e., the \sersic{} index $n = 1$) disk.  The left panel shows the comparison between the \sersic{} index $n$; the right panel shows the comparison between the effective radius $R_e$.  In each panel, symbols and lines are the same as assigned in Figure \ref{fig:goodness2sersic}.}
		\label{fig:expfit}
	\end{figure*}
	
	\subsection{The Morphology of the Disk}\label{Sec_Disc_Disk}
	
	\subsubsection{Exponential at Optical and NIR Bands}\label{Sec_Disc_Disk_OptNIR}
	
	In many studies involved with morphology of galaxies, surface-brightness profiles for galactic disks are commonly assumed with an exponential decline (i.e., the \sersic{} index $n = 1$) for simplicity.  By setting the \sersic{} index for the disk as a free parameter, our work shows that the parameter is almost around $n = 1$ in the wavelength range from optical to NIR (Table \ref{table_disk}, Figures \ref{fig:final} and \ref{fig:lowsn}), which corroborates the exponential form for the disk profile.  In particular, when fixing $n = 1$ for the disk at all wavebands (i.e., assuming an exponential disk) with the bulge still left in a free \sersic{} form, we find no significant difference in the result, except for the UV bands.
	
	Figure \ref{fig:expfit} shows comparisons between the same parameters for the one-\sersic{} plus one-exponential components and the two \sersic{} components.  After fixing the \sersic{} index for the disk into $n = 1$, the \sersic{} index for the bulge and the effective radius for both components at most wavebands are almost unchanged, but the effective radius for the disk at the four UV bands offsets towards larger values.  Due to the presence of the faint inner area plus the bright spiral arms in outer regions at UV bands, the \sersic{} index is small ($n$ = 0.1--0.3, i.e., the surface-brightness profile is flat).  If the \sersic{} index is artificially increased (to $n = 1$), then the effective radius will increase at the same time, in order to preserve the shape of the surface-brightness profile.
	
	\subsubsection{Flat at UV Bands}\label{Sec_Disc_Disk_UV}
	
\begin{figure*}[!ht]
		\centering
		\plotone{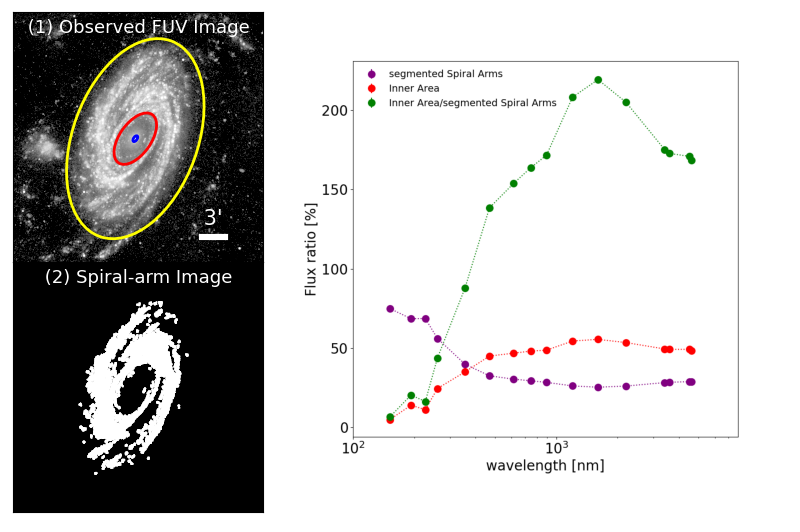}
		\caption{Left: the observed FUV image for M81 with the inner part of the galaxy marked as the area between the red and blue ellipses (the upper panel), where the yellow ellipse represents the aperture employed for photometry of the whole galaxy, and the spiral-arm image (the lower panel) segmented from the above FUV image.  Right: flux ratios of the spiral arms (purple) and the inner area (pink) to the whole galaxy as a function of wavelength, with the inner area to the spiral arms overplotted as well (green)}.
		\label{fig:spiral_flux_ratio}
	\end{figure*}

	From Figure \ref{fig:final} and Table \ref{table_disk}, we can see that, the disk at the UV bands appears to be extremely flat, with the \sersic{} index $n \sim 0.15$--0.33.  This flatness is attributed to a joint effect of the dim inner area and the bright spiral arms of the galaxy at the UV bands.  In this section, we inspect flux contributions of the inner area and the spiral arms to the whole galaxy at all of the wavebands, so as to infer the influence on the bulge-disk decomposition as a function of wavelength.  The approach and the result are displayed in Figure \ref{fig:spiral_flux_ratio}.
	
	Locations of the inner galactic area and the spiral arms are defined in the FUV image.  The inner area of the galaxy is enclosed in a manually determined elliptical aperture (marked in red) with the center R.A. = 148$^{\circ}$.888 and Dec. = 69$^{\circ}$.065 (consistent with the galactic center), the semi-major axis 212$\arcsec$, the ellipticity 0.47, and the PA $-35^{\circ}$.  The bulge-dominated innermost core, defined by a smaller elliptical aperture (marked in blue), is not included in the measurement of the inner area.  The aperture enclosing the core shares the center, the ellipticity, and the position angle the same with the larger aperture, but the semi-major axis is equal to two times the effective radius of the bulge at the FUV band.  The spiral arms are defined outside the inner area (red ellipse) and inside the aperture (yellow ellipse) for measuring the whole galaxy, by a threshold of the flux higher than 10 times the background uncertainty with more than 50 pixels.  The spiral arms are then segmented and loaded into a new image, as shown in the bottom-left panel of Figure \ref{fig:spiral_flux_ratio}.  This spiral-arm segmentation was applied to all of the images.
	
	The right panel of Figure \ref{fig:spiral_flux_ratio} shows the flux ratios of the inner area and the spiral arms to the whole galaxy as a function of wavelength.  From this diagram, we can see that, at the UV bands, the inner area trivially contributes $< 20\%$ flux to the whole galaxy, whereas the spiral arms predominantly account for $\sim 70\%$ and even more at the FUV band.  The situation is in contrast to that at the optical and NIR bands; more than half of the integrated flux for the galaxy is contributed by the inner area, but the percentage of the spiral arms drops to $\sim 30\%$.  As a result, the fitting of the radial surface-brightness profile for the disk inevitably results in a dimmer inner part and a brighter outer part at the UV bands than those at the optical and NIR bands, which causes the extreme flatness.
	
	The high flux fraction of the spiral arms found in the UV images are consistent with the picture of the density-wave theory (\citealt{Lin_1964}).  The spiral arms in M81 have been suggested to be a density-wave pattern (\citealt{Kendall_2008, Feng_2014, Wang_2015}).  When gas clouds go across spiral arms, this triggers large-scale shock, facilitating gravitational collapse, and production of new stars in gas clouds (\citealt{Roberts_1969, Seigar_2002, Kendall_2015, Yu_2018, Yu_2021, Pettitt_2020,Yu_2022b}).  The flux fraction of spiral arms decreases toward longer wavelengths where the emission is dominated by old stars.  Nevertheless, the flux fraction slightly increases beyond 3\,$\mu$m, probably caused by a contribution of polycyclic aromatic hydrocarbon (PAH) emission along the spiral arms (\citealt{Querejeta_2015}).
	
	\begin{figure*}[!ht]
		\centering
		\plotone{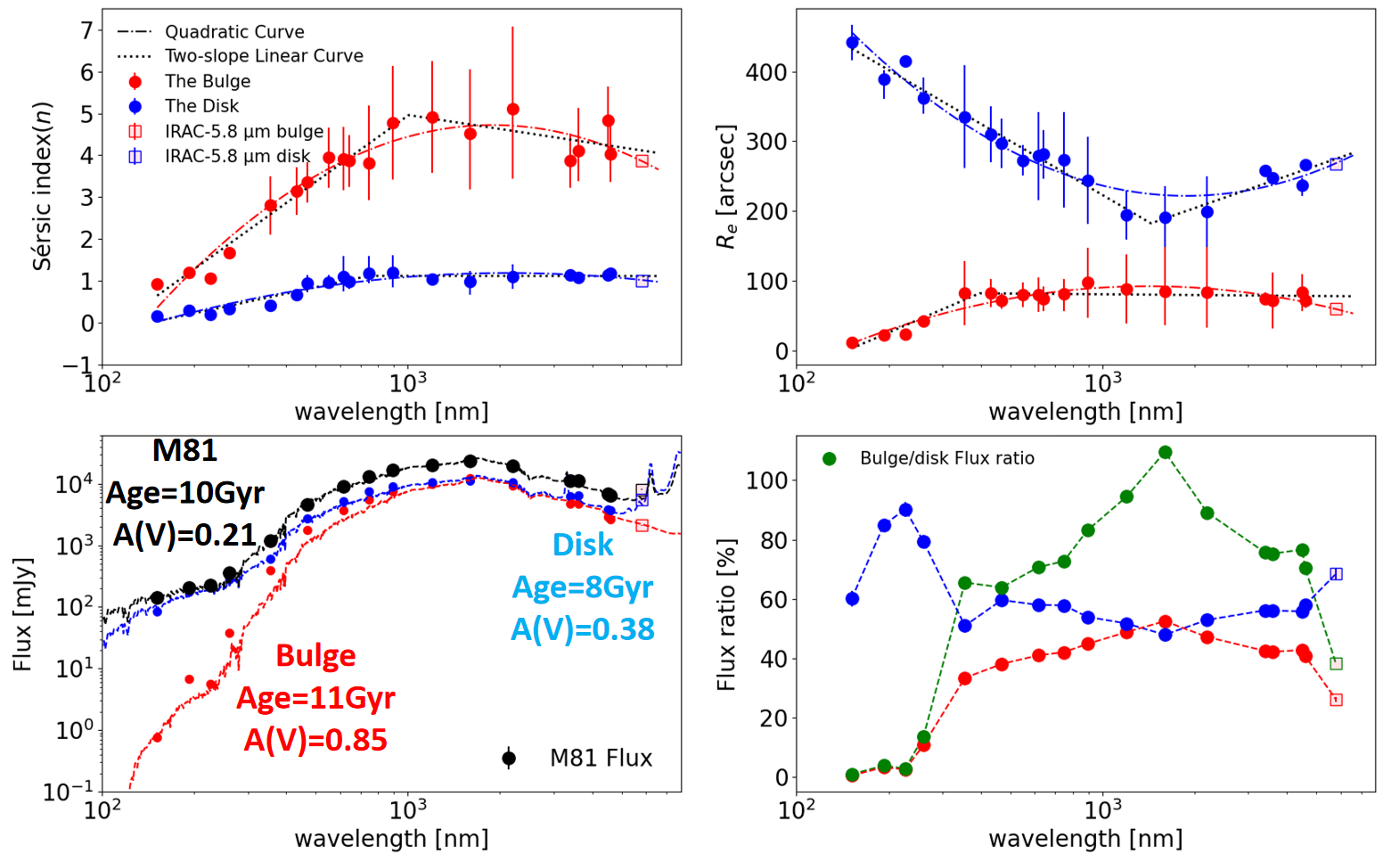}
		\caption{The same as Figure \ref{fig:final}, but the morphological parameters at the Spitzer-IRAC~5.8\,$\mu$m band are determined according to Equations (\ref{eq:n_bulge})--(\ref{eq:Re_disk}), so there is no error bar for the date points of $n$ and $R_e$ at this band.}
		\label{fig:finalnon}
	\end{figure*}
	
	\subsection{Practical Applicability of the Wavelength-dependent Morphology}\label{Sec_Disc_App}
	
	\begin{figure*}[!ht]
		\centering
		\plotone{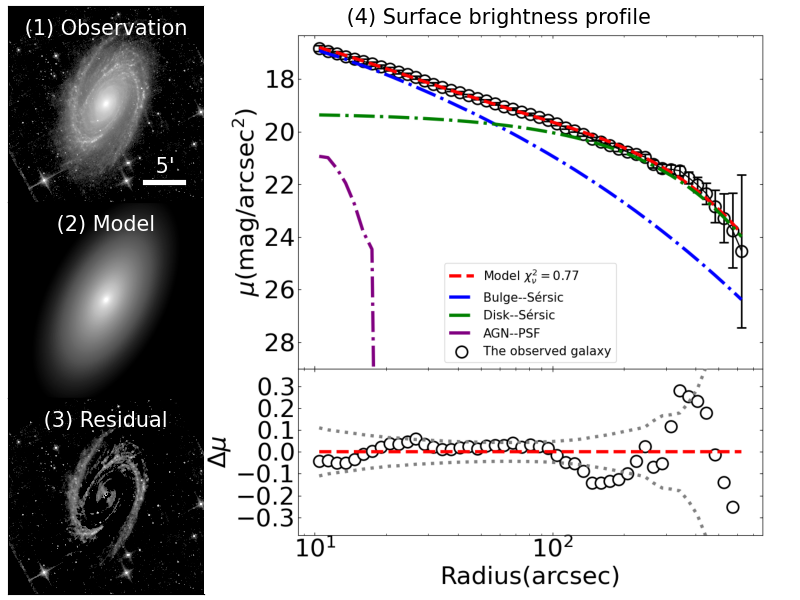}
		\caption{Results of the surface brightness fitting in the Spitzer-IRAC~5.8\,$\mu$m image.  The left panels show the observed (top panel), model (middle panel), and residual (bottom panel) images.  The top-right panel shows the radial surface brightness profiles of the observed galaxy and all of the components in the model, with the best-fit curves superimposed.  The bottom-right panel shows the comparison between the observed data and the model, and the gray dotted line indicates the error.  The legend is listed in the diagram.  During the decomposition, values for the \sersic{} index and the effective radius for the bulge and the disk in the model are determined according to Equations \ref{eq:n_bulge}--\ref{eq:Re_disk}.}
		\label{fig:renzao}
	\end{figure*}
	
	In the top panels of Figure \ref{fig:final}, the data points for \emph{Spitzer}-IRAC~5.8\,$\mu$m deviate from the best-fit curves except for the \sersic{} index for the bulge. We find that the inner area of M81 at the Spitzer-IRAC~5.8\,$\mu$m band contains dust emission in addition to starlight, which is a main cause of the deviation in the morphological parameters at this band.  Through our inspection, the distribution of the radiation of the internal dust is within 180$\arcsec$ of the radius.  These internal dust radiations mislead \GALFIT{} into allocating this part of light to the bulge model, which resulted in an overestimate of the effective radius and the total flux of the bulge. Because more starlight is allocated to the bulge, the brightness of the central area of the disk model decreases, resulting in an underestimate of the \sersic{} index for the disk, as discussed in Section \ref{Sec_Disc_Disk}.
	
	The wavelength-dependent morphology presented above provides a recipe for determining morphological parameters at any bands without radial surface-brightness fitting.  For M81 at the Spitzer-IRAC~5.8\,$\mu$m band, the radial surface-brightness fitting is affected by the presence of dust emission; as a consequence, the bulge-disk decomposition and the morphological estimate are conducted with considerable uncertainties.  Notwithstanding, by alternatively using Equations \ref{eq:n_bulge}--\ref{eq:Re_disk}, we are able to obtain the \sersic{} index and the effective radius for the bulge and the disk at 5.8\,$\mu$m, and subsequently create model images for the bulge and the disk at this band.
	
	Figures \ref{fig:finalnon} and \ref{fig:renzao} show results of the bulge-disk decomposition, with the \sersic{} index and the effective radius at 5.8\,$\mu$m determined by using the morphology-wavelength relation (i.e., Equations \ref{eq:n_bulge}--\ref{eq:Re_disk}) rather than obtained by fitting radial profiles of surface brightness.  This is an examination of the morphological estimate without radial surface-brightness fitting.  We can see that the newly created bulge and disk at 5.8\,$\mu$m are well fitted with the models in SEDs, but deviate from the main loci in flux ratios.  The integrated flux of the bulge decreases by 40.9\%, while the disk increases by 27.8\%.  In this case, the effect of the internal dust, which misled \GALFIT{} into allocating some part of the emission from the disk to the bulge, is eliminated.  Notwithstanding, the radial surface-brightness profiles for the newly created model have no obvious changes from the previous ones.

The relationship of morphological parameters with wavelength found in this work is usable for recovering galactic morphology at any band where bulge-disk decomposition and morphological fitting are not able to be carried out due to any reason.  According to our results, in the most general case, there is a simple approach to obtaining multiwavelength morphology of galaxies.  In this recipe, we suggest an initial estimate at the three characteristic wavebands $\lambda_\mathrm{1} < 3000$\,\AA, $\lambda_\mathrm{2} \sim 5500$--$6000$\,\AA, and $\lambda_\mathrm{3} > 1.0\,\mu$m (in the rest frame), and a subsequent interpolation at any other certain wavebands.

On the other hand, the results in this work bring forward a caveat that galaxies with classical bulges are likely misidentified for those with pseudo or no bulges if they are studied at rest-frame UV bands.  This seems especially crucial for high-redshift galaxies, which are usually observed through rest-frame UV channels with optical telescopes.  Our study of M81 is an attempt at quantifying morphological {\it K}-correction in a specific case.  Studies of additional galaxies of different types would be significant in acquiring comprehensive knowledge about wavelength-dependent morphology of galaxies, of which a wide range of applications can be made to a number of areas in practice.

	\begin{figure*}[!ht]
		\centering
		\plotone{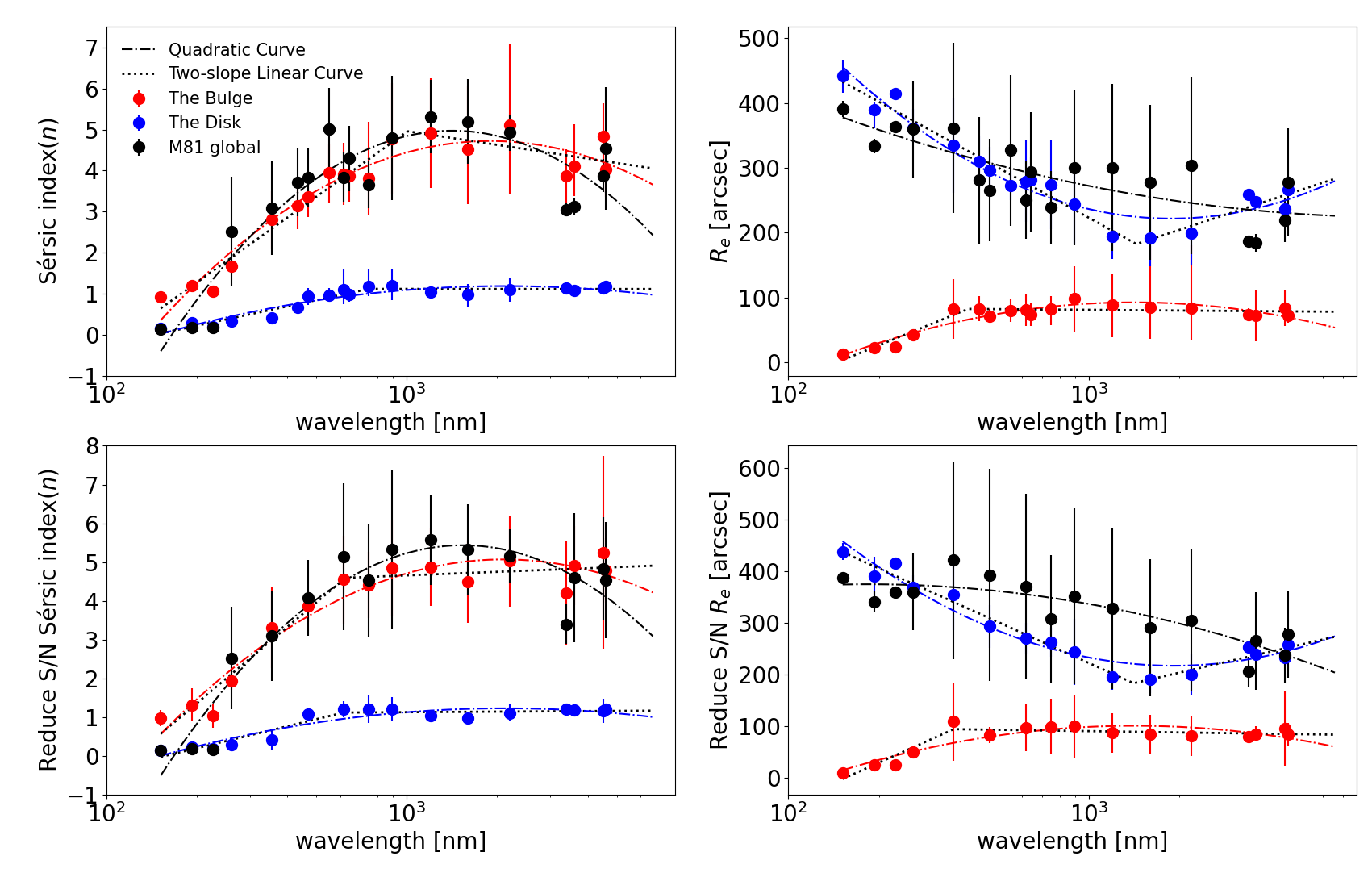}
		\caption{This figure presents the morphological parameters of the galaxy, including the \sersic{} index (top-left panel) and effective radius (top-right panel) fitting based on the original S/N data, as well as the \sersic{} index (lower-left panel) and effective radius (lower-right panel) fitting based on reduced S/N data. The black dots represent the fitting results for the integrated galaxy, which is a combination of all substructures.}
		\label{fig:final_and1}
	\end{figure*}
		
	\subsection{Comparison with Other Work}\label{Sec_Disc_Comp}
	
	Most previous studies involving multiwavelength morphology of galaxies have found very mild or even constant trends of the \sersic{} index with wavelength.  In general cases, galaxies have been found to be more compact and smaller at longer wavelengths.  For early-type galaxies or classical bulges observed through optical--NIR channels (approximately from u to K$_s$ bands), the \sersic{} index increases with wavelength, but the variations are shallow or flat ($n \sim 3$--$4$, \citealt{Kelvin_2012, Vika_2013, Vulcani_2014, Rebecca_2015, Victor_2018, Lima-Dias_2021}); in the same wavelength range, late-type galaxies or disks appears invariable in concentration level (commonly with the \sersic{} index $n \sim 1$, \citealt{Kelvin_2012, Vulcani_2014, Lumbreras_2019, Lima-Dias_2021}).  When observed at UV bands ($\lambda < 3000$\,\AA), early-type galaxies or classical bulges appear with lower values for the \sersic{} index ($n \sim 2$--$3$, \citealt{Rampazzo_2017, Leahy_2023}).  Nonetheless, morphology of galaxies and galactic subsystems at wavelength extending to FUV or beyond K$_s$ bands has not yet been well explored in combination with optical--NIR channels.  By collecting 20-band imaging data spanning from 1516\,\AA~to 5.8 $\mu$m, our study creates the widest wavelength window into quantitative galactic morphology to date.  According to our results, if the wavelength is confined in the optical--NIR range, particularly 5500\,\AA--1.0\,$\mu$m, then the morphology appears to be virtually changeless with wavelength.
	
	Most of the previous studies on this topic have measured galaxies as a mixture of all substructures, while in this work we carry out the bulge-disk decomposition and separately measure the individual substructures, which is an important difference of our study from others.  In order to examine the effect of the bulge-disk or no decomposition on the resultant morphology as a function of wavelength, we perform morphological measurement with no bulge-disk decomposition, i.e., using a single \sersic{} component (plus a PSF component to fit the AGN in M81) to fit the surface-brightness profiles at all wavebands.  The results, with both original and equalized S/N, are displayed in Figure \ref{fig:final_and1}.  It is clearly displayed that, for the galaxy as a mixture with no decomposition, the \sersic{} index is dominated by the disk at the UV bands and by the bulge at the optical--NIR bands; the effective radius is determined solely by the disk, but decreases monotonically with wavelength; both trends remain to be obvious but show large dispersion.  By comparison between the results of the two- and one-\sersic{} fits, we find that the one-\sersic{} fit yields a roughly consistent estimate of the \sersic{} index with that for the bulge at optical and NIR bands, albeit an overestimates of the effective radius.  However, if the wavelength is confined to a narrow range, particularly at $\lambda \leq 2\,\mu$m, the variation in morphology with wavelength appears to be monotonic with large dispersion.
	
	In practice, quite a large number of investigations have assumed constant morphology at different wavelengths for simplicity (\citealt{McDonald_2011, Marina_2014, Rebecca_2016, Buzzo_2021}).  According to our work, this assumption is generally safe at the rest-frame wavelength $\lambda \geq 5500$\,\AA, but it takes a high risk of overestimating the \sersic{} index for both of the bulge and the disk at shorter wavelengths, particularly when $\lambda < 3000$\,\AA~(in the rest frame).
	
	\section{\textbf{SUMMARY}}\label{Sec_Sum}
	
	In this work, we collect imaging data at 20 wavebands in total from FUV to NIR obtained with space and ground-based telescopes/surveys, including \emph{GALEX}, \emph{Swift}, WIYN, SDSS, 2MASS, \emph{WISE}, and \emph{Spitzer}, to investigate multiwavelength morphology of the nearby early-type spiral galaxy M81.  By fitting radial surface brightness in the images, we carry out bulge-disk decomposition and derive a series of morphological parameters as a function of wavelength.  We find that, the \sersic{} index for the bulge and the effective radius for the disk are tightly correlated with wavelength, where the \sersic{} index for the bulge increases sharply from $n \sim 1$ to $n \sim 4$ from UV to optical bands and varies smoothly between $n \sim 4$ and $5$ at the optical--NIR bands, while the effective radius for the disk emerges with a similar intensity of the variation but in an opposite direction.
	
	It is worth noting that the bulge in M81 is classical in type, with the \sersic{} index $n \sim 4$--$5$ at optical and NIR bands, but it behaves like a pseudo-bulge or no bulge at UV bands with $n \sim 1$.  This hints that, observing through UV channels, we are likely to mistake a classical bulge for a pseudo-bulge or no bulge.  It is particularly important for high-redshift galaxies, of which morphology is usually studied at rest-frame UV bands.  The relationship of morphological parameters with wavelength discovered in this work provides panchromatic insight into the morphological {\it K}-correction and is usable for recovering the galactic morphology at any band where morphological measurements are not able to be carried out.
	
	Previous investigations on multi-band morphology of galaxies were often confined in the optical range with no separation of substructures, and they found a monotonic and smooth change, or even an invariable trend, in morphology with wavelength.  With data spanning a wider wavelength range collected and the bulge-disk composition performed, the morphological characteristics revealed in this work are more comprehensive and convincing.  On the basis of the multiwavelength morphology, we plan to study spatially resolved SEDs for the bulge and the disk in M81 in the next work of this series, aimed at exploring stellar population properties and star formation/quenching history for the galaxy and the galactic subsystems.


	\section*{\textbf{ACKNOWLEDGMENTS}}\label{ACKNOWLEDGMENTS}
	
	We are extremely grateful to Luis C. Ho for constructive comments and discussion.  We appreciate the careful reading and the insightful comments offered by the anonymous reviewer, which have helped us to improve the paper significantly.  This work is supported by the National Natural Science Foundation of China (NSFC, Nos. U2031106 and U2031201).   Jun-Yu Gong especially thanks Luis C. Ho, Ming-Yang Zhuang, and Ruan-Cun Li for their help and hospitality during the visit to the Kavli Institute for Astronomy and Astrophysics at Peking University (KIAA-PKU).  Si-Yue Yu acknowledges the support of the Alexander von Humboldt Foundation.  Kavli IPMU was established by World Premier International Research Center Initiative (WPI), MEXT, Japan.

This research is based on observations made with the \emph{Galaxy Evolution Explorer}, and data obtained from the Mikulski Archive for Space Telescopes (MAST) at the Space Telescope Science Institute, which is operated by the Association of Universities for Research in Astronomy, Inc., under NASA contract NAS 5-26555.  This research has made use of data provided by the High Energy Astrophysics Science Archive Research Center (HEASARC), which is a service of the Astrophysics Science Division at NASA/GSFC.  This research has made use of data from SDSS-III, which is funded by the Alfred P. Sloan Foundation, the Participating Institutions, the National Science Foundation, and the U.S. Department of Energy Office of Science; the SDSS-III web site is \url{http://www.sdss3.org/}; SDSS-III is managed by the Astrophysical Research Consortium for the Participating Institutions of the SDSS-III Collaboration including the University of Arizona, the Brazilian Participation Group, Brookhaven National Laboratory, Carnegie Mellon University, University of Florida, the French Participation Group, the German Participation Group, Harvard University, the Instituto de Astrofisica de Canarias, the Michigan State/Notre Dame/JINA Participation Group, Johns Hopkins University, Lawrence Berkeley National Laboratory, Max Planck Institute for Astrophysics, Max Planck Institute for Extraterrestrial Physics, New Mexico State University, New York University, Ohio State University, Pennsylvania State University, University of Portsmouth, Princeton University, the Spanish Participation Group, University of Tokyo, University of Utah, Vanderbilt University, University of Virginia, University of Washington, and Yale University.  This research has made use of the NASA/IPAC Extragalactic Database (NED) which is funded by the National Aeronautics and Space Administration and operated by the California Institute of Technology.  This research has made use of the NASA/IPAC Infrared Science Archive (IRSA), which is funded by the National Aeronautics and Space Administration and operated by the California Institute of Technology.  This research has also made use of the SAO/NASA's Astrophysics Data System (ADS) Bibliographic Services.
	

        \appendix
	
	\renewcommand\thefigure{\Alph{section}\arabic{figure}}
	\setcounter{figure}{0}
	\renewcommand\thetable{\Alph{section}\arabic{table}}
	\setcounter{table}{0}
	
	\section{\textbf{Products of the Multiwavelength Bulge-Disk Decomposition for M81}}\label{Sec_Appendix_A}

	\vspace*{+2mm}

	In this appendix, we present products of the Bulge-Disk decomposition at the 20 wavebands (the process of the decomposition is presented in Section \ref{Sec_Decomp} in detail).   Morphological and flux parameters for the bulge and the disk are listed in Tables \ref{table_bulge} and \ref{table_disk}, respectively.
	
\begin{figure}
		\figurenum{A1}
		\plotone{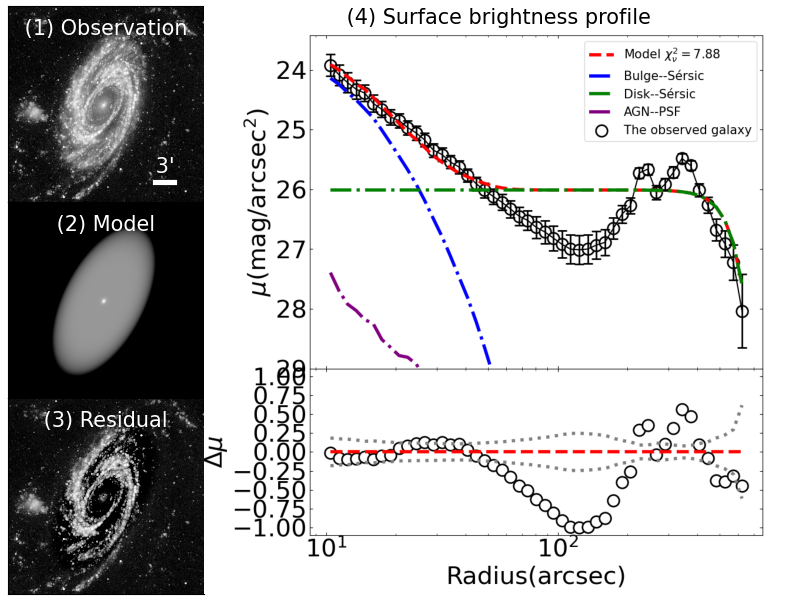}
		\caption{Results of the surface brightness fitting in the \emph{GALEX} FUV image.  The left panels show the observed (top panel), modeled (middle panel), and residual (bottom panel) images.  The top-right panels show the radial surface brightness profiles of the observed galaxy and all of the components in the model, with the best-fit curves superimposed.  The bottom-right panel show the comparison between the observed data and the model, and the gray dotted line indicates the error.  The legend is listed in the diagram.}
		\label{fig:fd}
	\end{figure}

\begin{figure}
		\figurenum{A2}
		\plotone{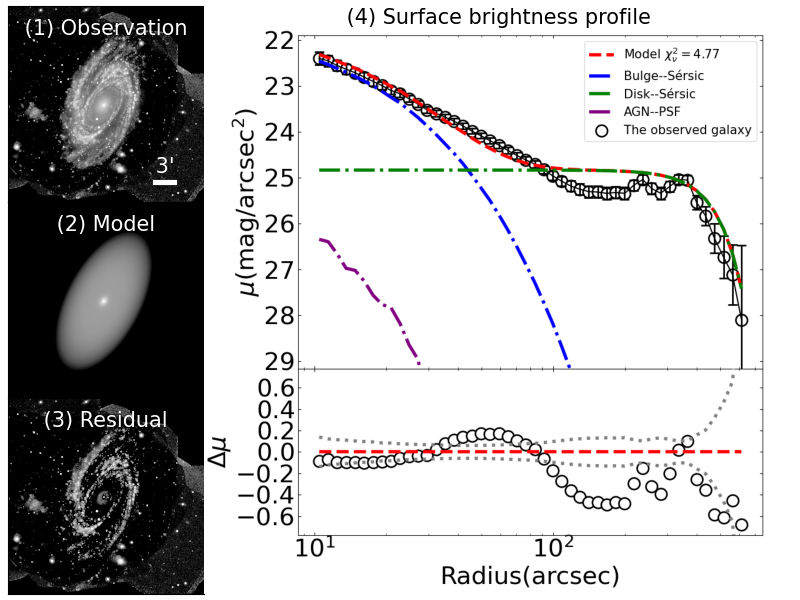}
		\caption{The same as Figure \ref{fig:fd} but in the \emph{Swift} UVW2 image.}
	 \label{fig:uvw2}
	\end{figure}

\begin{figure}
		\figurenum{A3}
		\plotone{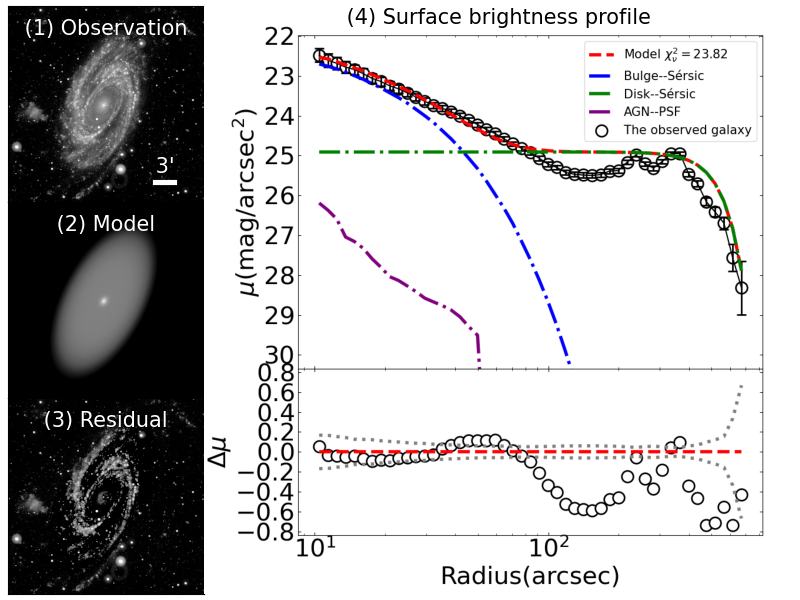}
		\caption{The same as Figure \ref{fig:fd} but in the \emph{GALEX} NUV image.}
	 \label{fig:nd}
	\end{figure}

\begin{figure}
		\figurenum{A4}
		\plotone{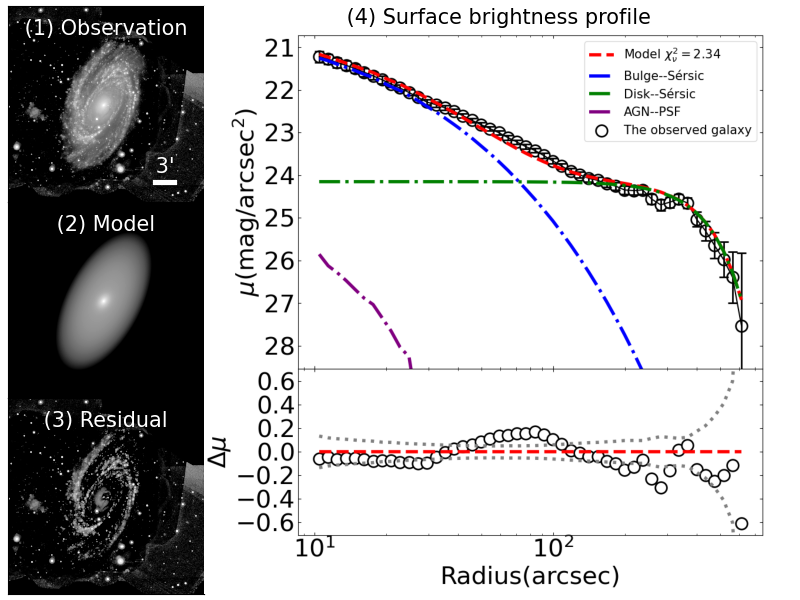}
		\caption{The same as Figure \ref{fig:fd} but in the \emph{Swift} UVW1 image.}
	 \label{fig:uvw1}
	\end{figure}

\begin{figure}
		\figurenum{A5}
		\plotone{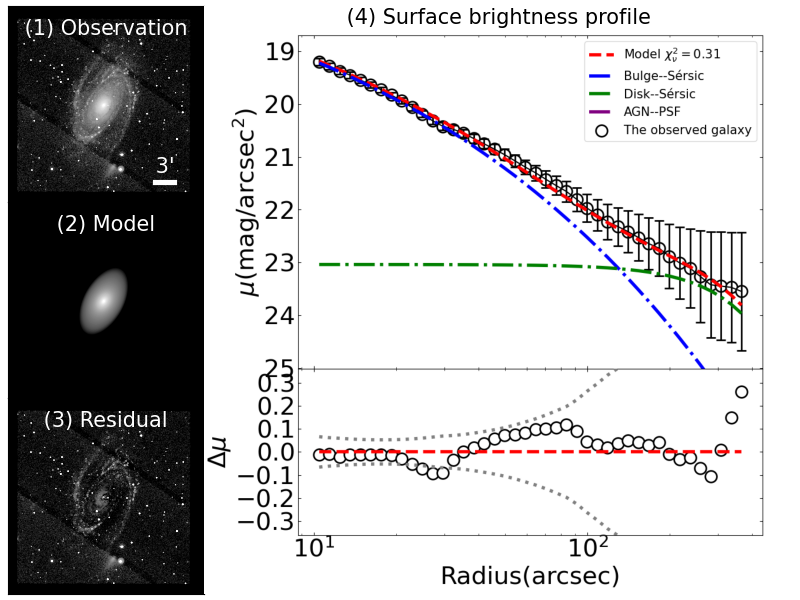}
		\caption{The same as Figure \ref{fig:fd} but in the SDSS {\it u} image.}
	 \label{fig:u}
	\end{figure}

\begin{figure}
		\figurenum{A6}
		\plotone{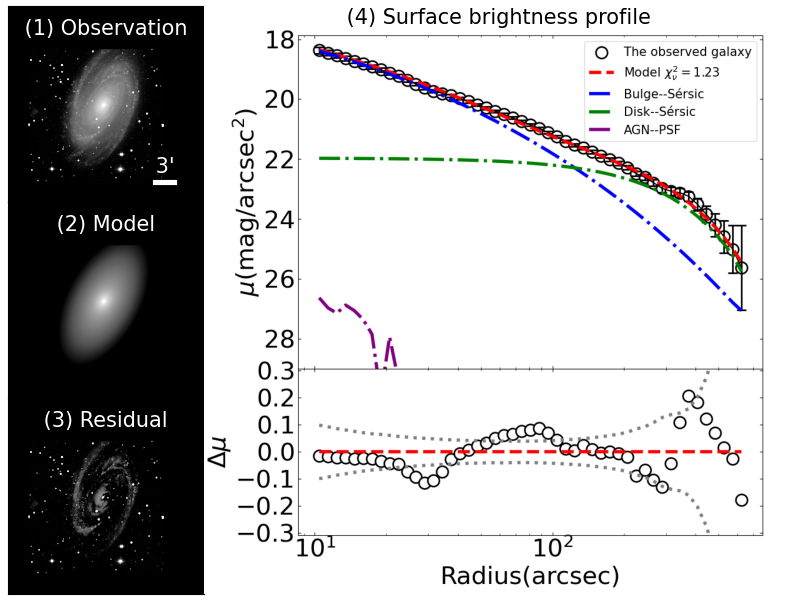}
		\caption{The same as Figure \ref{fig:fd} but in the WIYN B image.}
	 \label{fig:B}
	\end{figure}

\begin{figure}
		\figurenum{A7}
		\plotone{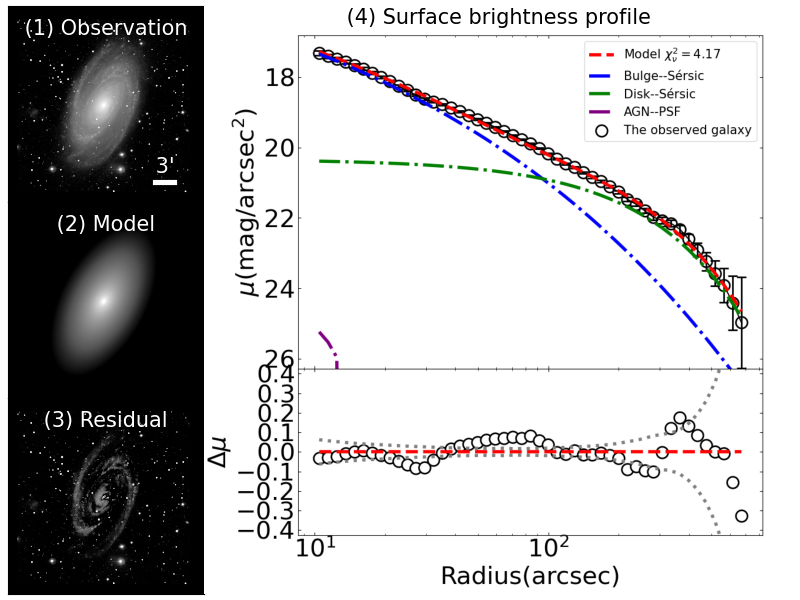}
		\caption{The same as Figure \ref{fig:fd} but in the SDSS {\it g} image.}
	 \label{fig:g}
	\end{figure}

\begin{figure}
		\figurenum{A8}
		\plotone{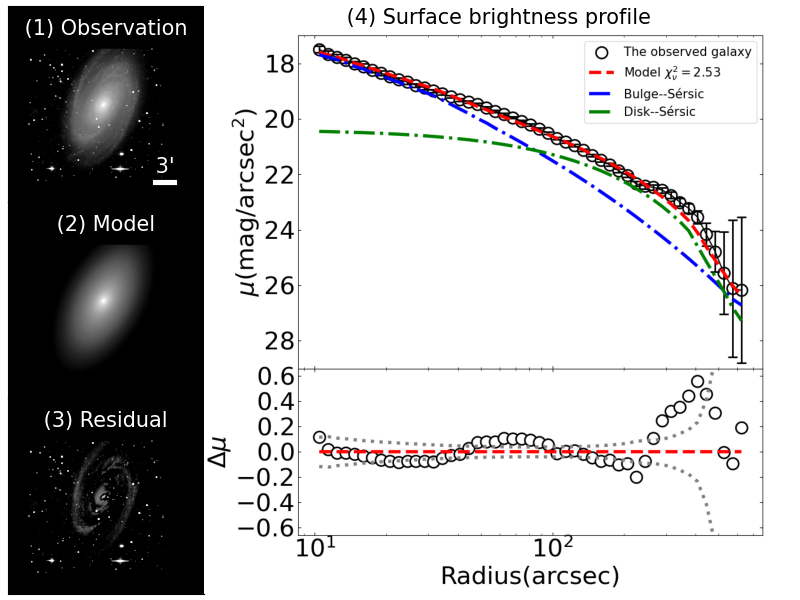}
		\caption{The same as Figure \ref{fig:fd} but in the WIYN V image.}
	 \label{fig:V}
	\end{figure}

\begin{figure}
		\figurenum{A9}
		\plotone{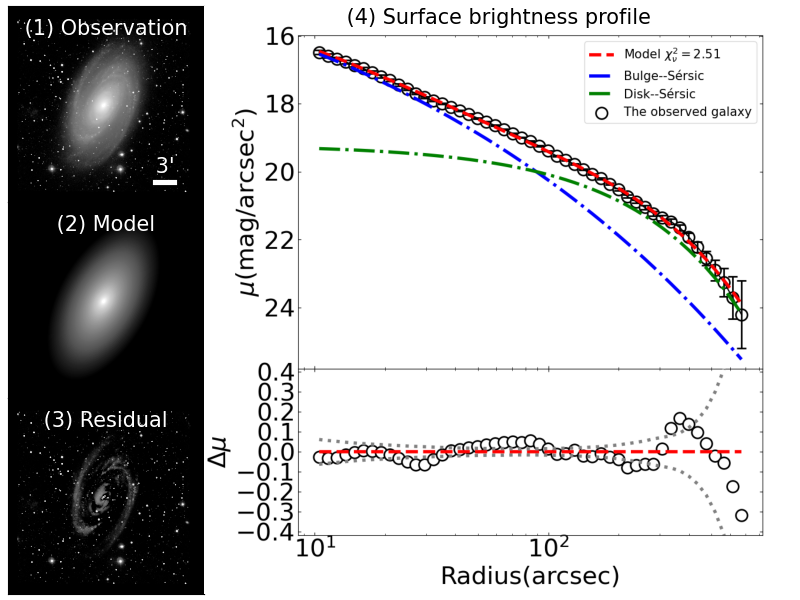}
		\caption{The same as Figure \ref{fig:fd} but in the SDSS {\it r} image.}
	 \label{fig:sdssr}
	\end{figure}

\begin{figure}
		\figurenum{A10}
		\plotone{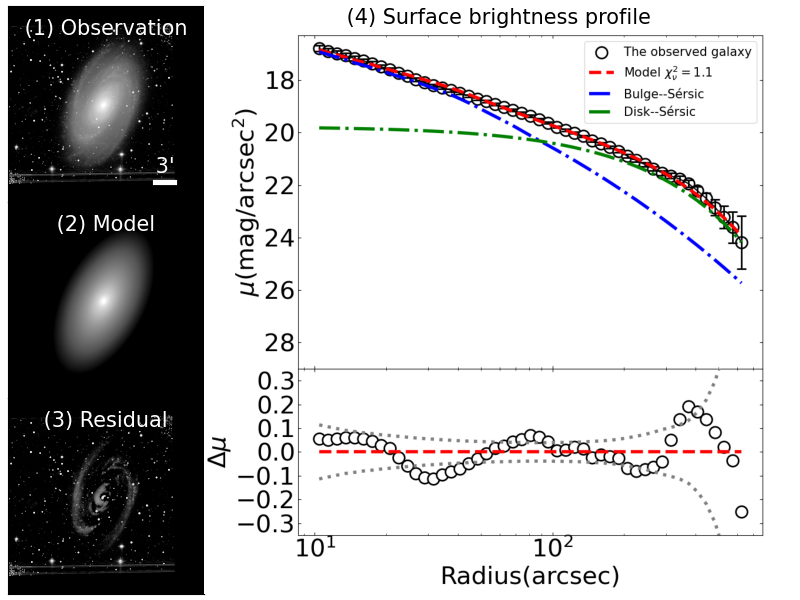}
		\caption{The same as Figure \ref{fig:fd} but in the WIYN R image.}
	 \label{fig:r}
	\end{figure}

\begin{figure}
		\figurenum{A11}
		\plotone{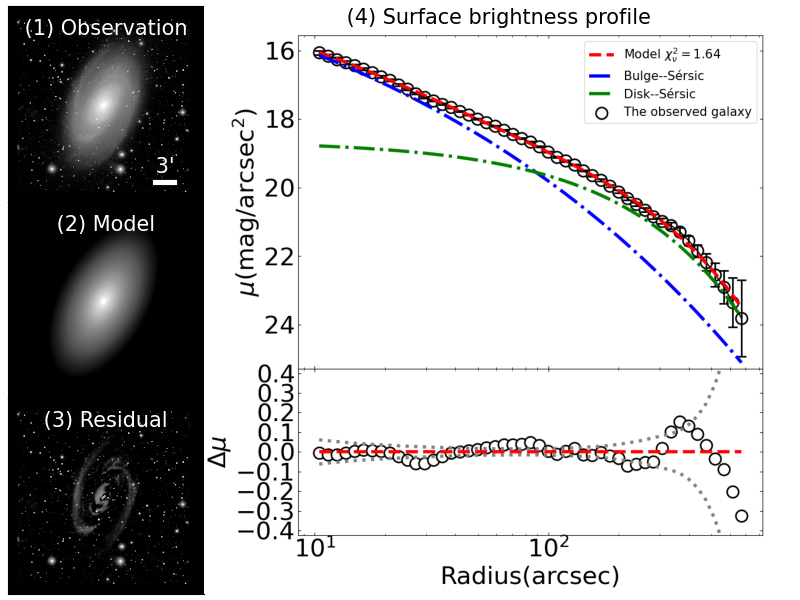}
		\caption{The same as Figure \ref{fig:fd} but in the SDSS {\it i} image.}
	 \label{fig:i}
	\end{figure}

\begin{figure}
		\figurenum{A12}
		\plotone{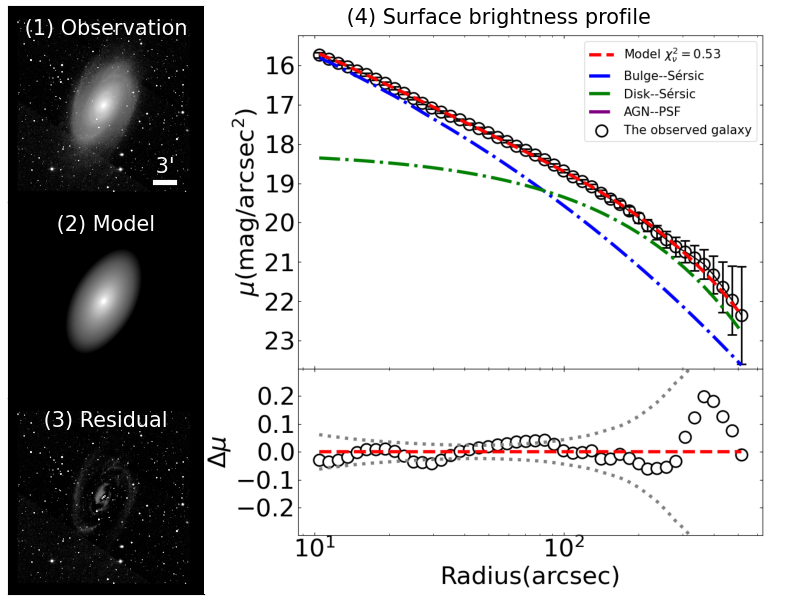}
		\caption{The same as Figure \ref{fig:fd} but in the SDSS {\it z} image.}
	 \label{fig:z}
	\end{figure}

\begin{figure}
		\figurenum{A13}
		\plotone{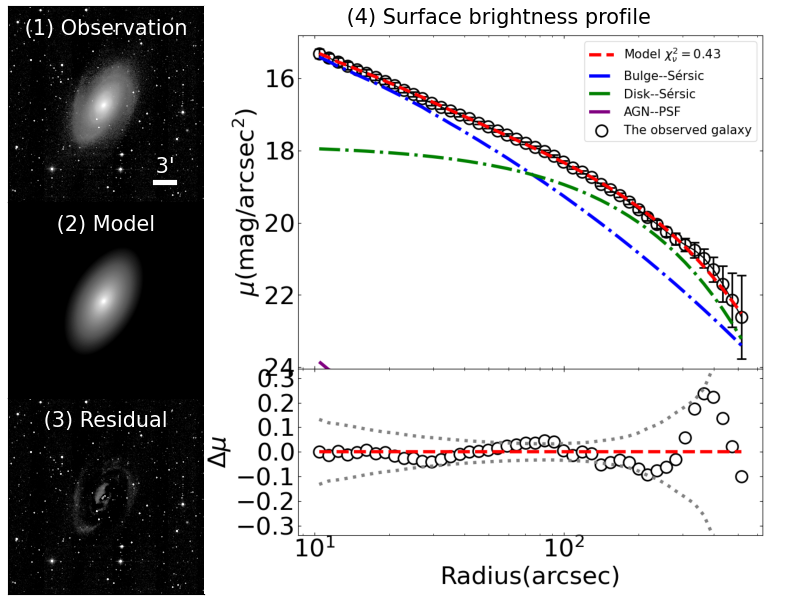}
		\caption{The same as Figure \ref{fig:fd} but in the 2MASS J image.}
	 \label{fig:j}
	\end{figure}

\begin{figure}
		\figurenum{A14}
		\plotone{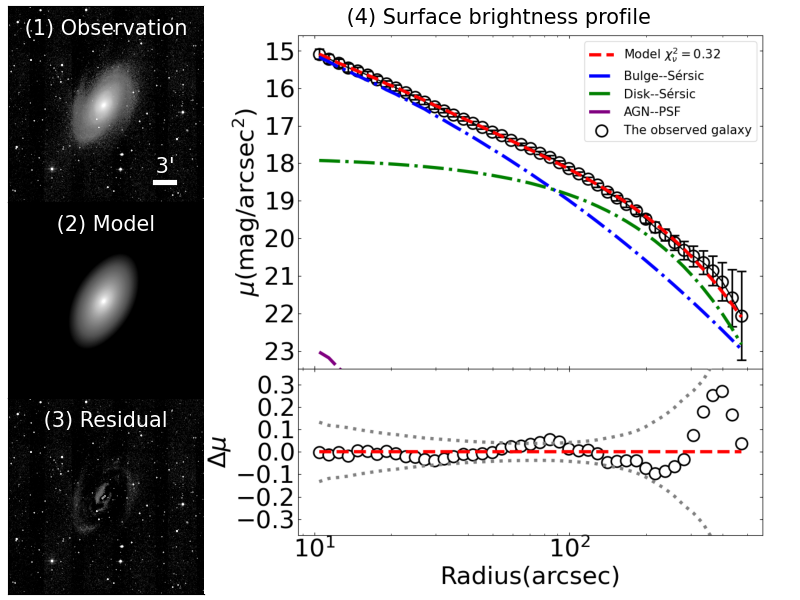}
		\caption{The same as Figure \ref{fig:fd} but in the 2MASS H image.}
	 \label{fig:h}
	\end{figure}

\begin{figure}
		\figurenum{A15}
		\plotone{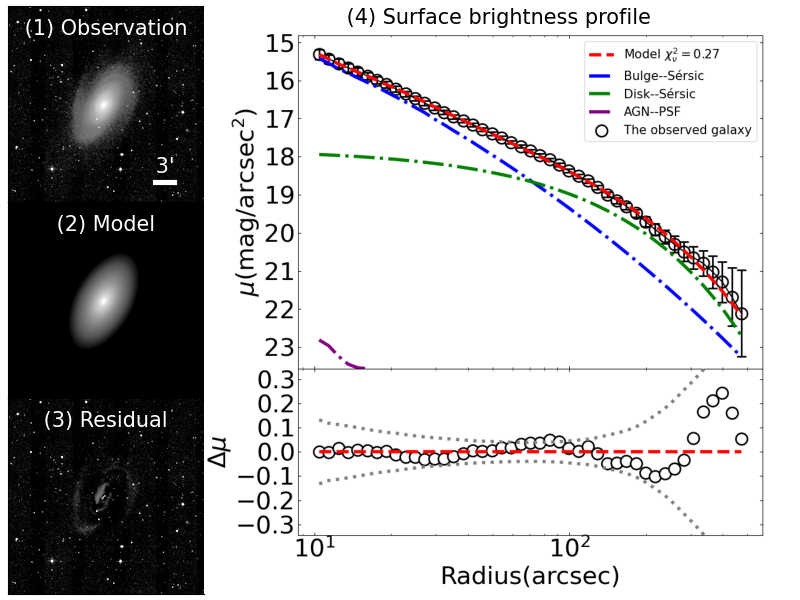}
		\caption{The same as Figure \ref{fig:fd} but in the 2MASS K$_s$ image.}
	 \label{fig:k}
	\end{figure}

\begin{figure}
		\figurenum{A16}
		\plotone{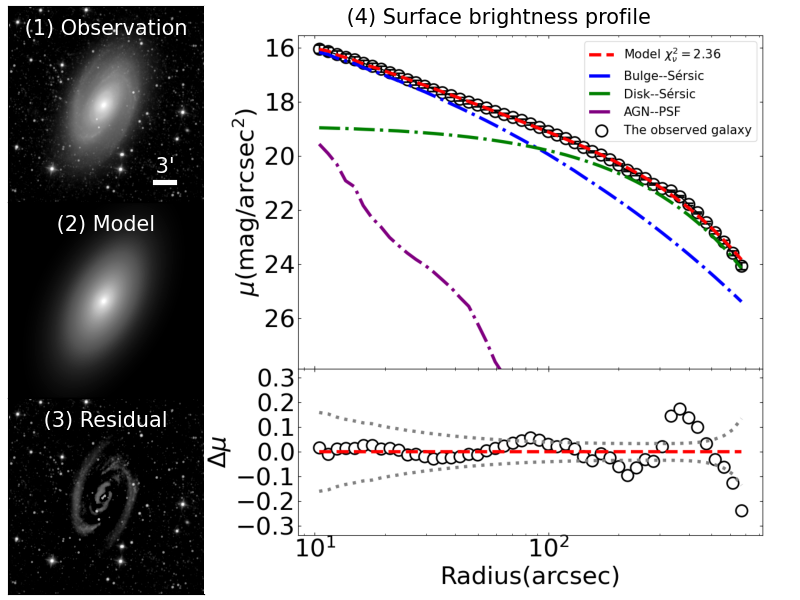}
		\caption{The same as Figure \ref{fig:fd} but in the \emph{WISE} W1 image.}
 	 \label{fig:W1}
	\end{figure}

\begin{figure}
		\figurenum{A17}
		\plotone{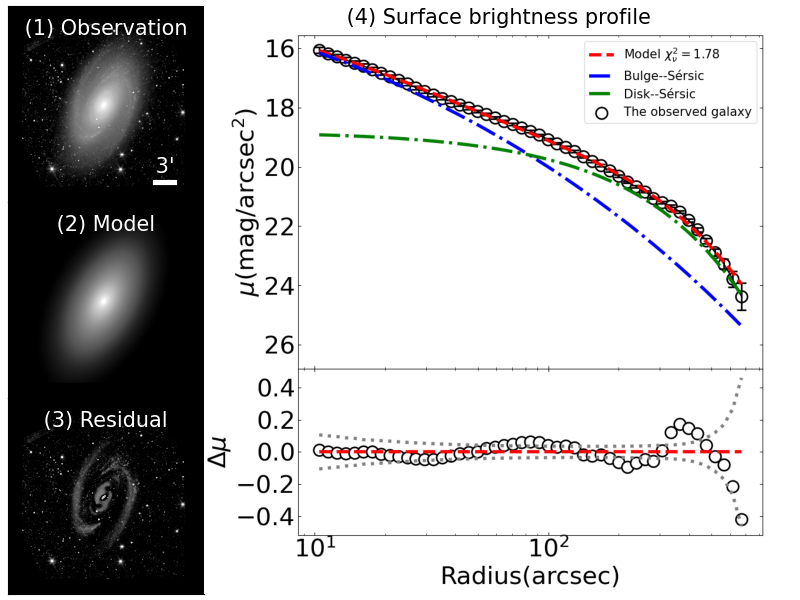}
		\caption{The same as Figure \ref{fig:fd} but in the \emph{Spitzer}-IRAC I1 image.}
	 \label{fig:I1}
	\end{figure}

\begin{figure}
		\figurenum{A18}
		\plotone{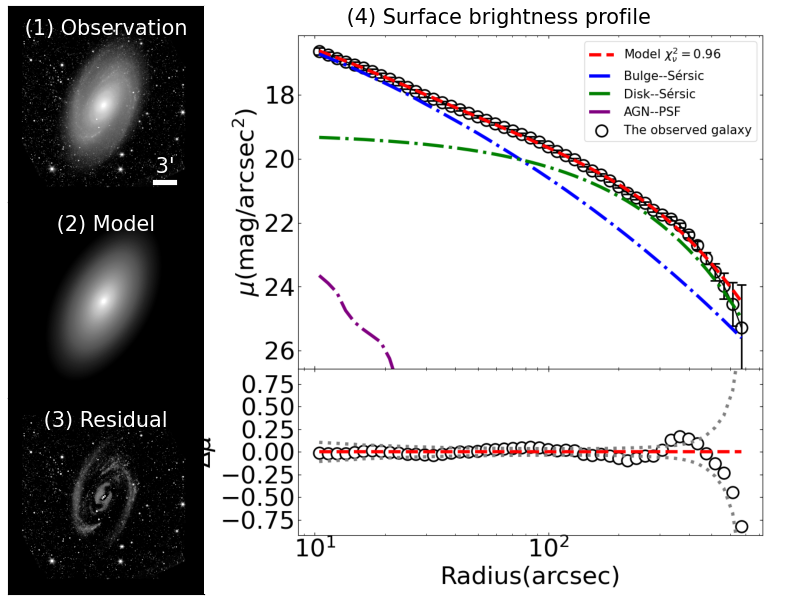}
		\caption{The same as Figure \ref{fig:fd} but in the \emph{Spitzer}-IRAC I2 image.}
 	\label{fig:I2}
	\end{figure}

\begin{figure}
		\figurenum{A19}
		\plotone{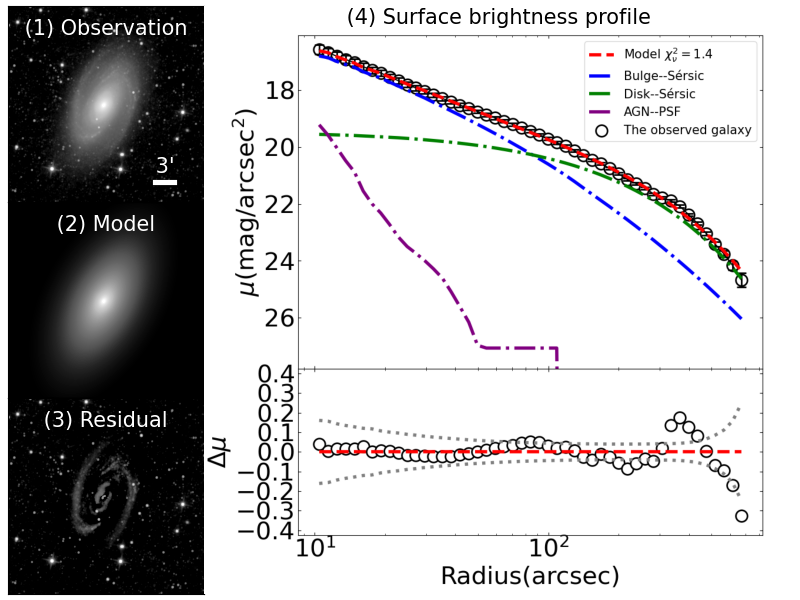}
		\caption{The same as Figure \ref{fig:fd} but in the \emph{WISE} W2 image.}
	 \label{fig:W2}
	\end{figure}

\begin{figure}
		\figurenum{A20}
		\plotone{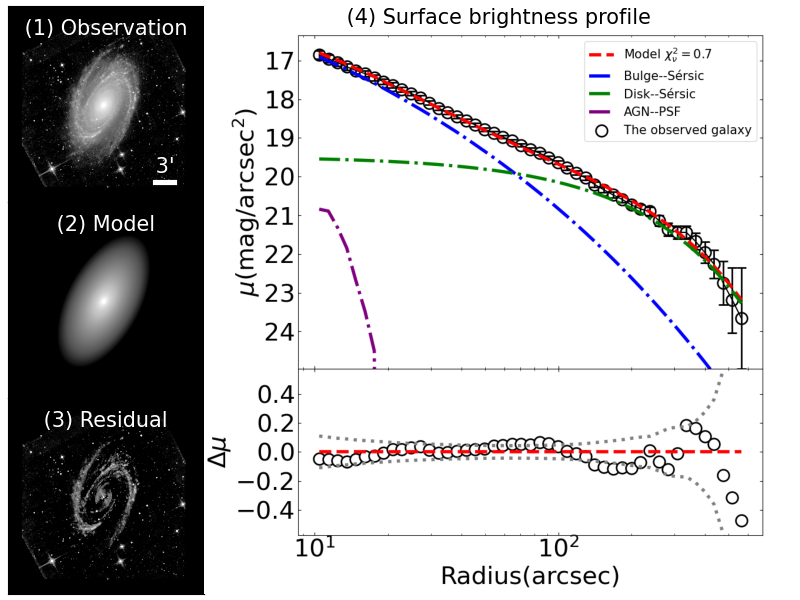}
		\caption{The same as Figure \ref{fig:fd} but in the \emph{Spitzer}-IRAC I3 image.}
	 \label{fig:I3}
	\end{figure}

	Figures \ref{fig:fd}--\ref{fig:I3} show observed, modeled, and residual images, as well as radial profiles of surface brightness for the bulge, the disk, and the whole galaxy, at the 20 wavebands.  The model is a sum of all morphological fitting components in the decomposition at a certain waveband.  For most of the wavebands, the fitting components include the bulge, the disk, and the PSF component (representing the AGN), while for SDSS-{\it r}, SDSS-{\it i}, WIYN-V, WIYN-R, and Spitzer-IRAC~3.6\,$\mu$m, the M81 center is affected by saturated pixels, so that we do not take the PSF component into the decomposition.  The residual image is obtained by subtracting the model image from the observed image.  The radial profiles of surface brightness are produced by using the same method as in \citet{Gao_2017}.  First, we need to obtain the ellipticity, the position angle, and the central coordinates of M81. We separately carry out isophote fitting for images of different bands, and we use the isophote for which brightness reaches the background uncertainty as the aperture of that band. We define the ellipticity, position angle and center coordinates of M81 as the average of the aperture properties in each band. We create multiple ellipses with the same central coordinates, axis ratio, and position angle as M81, but with a regular change in size. The median flux for the area between two adjacent ellipses is counted and converted into AB magnitude. When the median flux at an outer area reaches the background uncertainty, we set the area as the outermost part of the radial profiles.  At a certain waveband, the uncertainty in the radial profile is an orthogonal combination of the background uncertainty, the calibration error, and the isophote fitting error. We consider the error of isophote here because the ellipticity, the position angle, and the central coordinates of M81 all come from the method of isophote fitting.
	
	A reduced chi square ($\chi^2_{\nu}$) is employed as a quantification of the goodness of the fit and defined as follows:
	
	\begin{equation}\label{eq:chi2}
		\chi_{\nu}^{2}=\frac{1}{N_{\mathrm{dof}}}\sum_{i=1}^{N_{\mathrm{data}}}\frac{\left(m_{\mathrm{data},i}-m_{\mathrm{model},i}\right)^{2}}{\sigma_{m,i}^{2}} ~,
	\end{equation}
	
	where $N_{\mathrm{dof}}$ is the degrees of freedom; $N_{\mathrm{data}}$ is the number of data points in the surface brightness profile; $m_{\mathrm{data}}$ and $m_{\mathrm{model}}$ represent the surface brightness (in units of magnitude) for data and model at a certain radius, respectively; $\sigma_m$ is the error in the data.
	
	\newpage
	
	\begin{center}
		\begin{table}[h]
			\centering
			\scriptsize
			\caption{Multiwavelength Statistics of the Bulge as a Result of the Decomposition}
			\label{table_bulge}
			\setlength{\tabcolsep}{0.1mm}{
				\begin{tabular}{ccrrrrr}
					\toprule{}
					\multirow{1}{*}{Telescope} & \multirow{1}{*}{Filter} & \multirow{1}{*}{\begin{tabular}[c]{@{}c@{}}Wavelength\\ \\ {(}\AA{)}\end{tabular}} & \multirow{1}{*}{\begin{tabular}[c]{@{}c@{}}\sersic{} index \\ ($n$)\end{tabular}} & \multirow{1}{*}{\begin{tabular}[c]{@{}c@{}}Effective radius\\ ($R_e$) \\ {(}arcsec{)}\end{tabular}} & \multirow{1}{*}{\begin{tabular}[c]{@{}c@{}}Flux denisity\\ \\ {(}mJy{)}\end{tabular}} & \multirow{1}{*}{\begin{tabular}[c]{@{}c@{}}Flux ratio\\ \\ {(}\%{)}\end{tabular}} \\
					&                         &                                                                               &                                                                                                  &                                                                                              &                                                                                    &                                                                                \\
					&                         &                                                                               &                                                                                                  &                                                                                              &                                                                                    &                                                                                \\ 						\midrule{}
					\emph{GALEX}                      & FUV                     & 1516                                                                          & 0.92$^{+0.01}_{-0.01}$                                                                          & 12.15$^{+0.24}_{-0.09}$                                                                      & 0.75$\pm$0.02                                                                      & 0.53$\pm$0.02                                                                           \\
					\emph{Swift}                      & UVW2                    & 1928                                                                          & 1.19$^{+0.01}_{-0.01}$                                                                          & 22.99$^{+0.17}_{-0.09}$                                                                      & 6.85$\pm$0.09                                                                      & 3.29$\pm$0.09                                                                           \\
					\emph{GALEX}                      & NUV                     & 2267                                                                          & 1.06$^{+0.01}_{-0.01}$                                                                          & 23.66$^{+0.08}_{-0.05}$                                                                      & 5.69$\pm$0.08                                                                      & 2.53$\pm$0.07                                                                           \\
					\emph{Swift}                      & UVW1                    & 2600                                                                          & 1.66$^{+0.04}_{-0.02}$                                                                          & 43.02$^{+1.81}_{-1.07}$                                                                      & 37.84$\pm$0.51                                                                     & 10.73$\pm$0.29                                                                          \\
					SDSS                       & {\it u}                       & 3551                                                                          & 2.80$^{+0.70}_{-0.70}$                                                                          & 82.40$^{+45.94}_{-45.80}$                                                                    & 397.23$\pm$1.62                                                                    & 33.37$\pm$0.27                                                                          \\
					WIYN                       & B                       & 4331                                                                          & 3.14$^{+0.57}_{-0.57}$                                                                          & 82.61$^{+18.69}_{-18.70}$                                                                    & ---                                                                               & ---                                                                           \\
					SDSS                       & {\it g}                       & 4686                                                                          & 3.36$^{+0.48}_{-0.48}$                                                                          & 71.62$^{+10.60}_{-10.60}$                                                                    & 1747.31$\pm$6.99                                                                   & 38.09$\pm$0.30                                                                          \\
					WIYN                       & V                       & 5500                                                                          & 3.94$^{+0.71}_{-0.72}$                                                                          & 80.19$^{+16.88}_{-16.89}$                                                                    & ---                                                                               & ---                                                                           \\
					SDSS                       & {\it r}                       & 6166                                                                          & 3.92$^{+0.75}_{-0.75}$                                                                          & 80.58$^{+24.28}_{-24.46}$                                                                    & 3706.65$\pm$14.83                                                                  & 40.98$\pm$0.33                                                                          \\
					WIYN                       & R                       & 6425                                                                          & 3.84$^{+0.62}_{-0.62}$                                                                          & 74.15$^{+17.83}_{-17.84}$                                                                    & ---                                                                               & ---                                                                           \\
					SDSS                       & {\it i}                       & 7480                                                                          & 3.80$^{+1.38}_{-0.87}$                                                                          & 81.91$^{+20.75}_{-24.98}$                                                                    & 5476.54$\pm$19.17                                                                  & 42.05$\pm$0.29                                                                          \\
					SDSS                       & {\it z}                       & 8932                                                                          & 4.78$^{+1.35}_{-1.35}$                                                                          & 97.88$^{+50.13}_{-50.13}$                                                                    & 7410.17$\pm$29.66                                                                  & 44.83$\pm$0.36                                                                          \\
					2MASS                      & J                       & 12000                                                                         & 4.91$^{+1.33}_{-1.33}$                                                                          & 88.15$^{+49.49}_{-48.49}$                                                                    & 9928.89$\pm$84.41                                                                  & 48.84$\pm$0.83                                                                          \\
					2MASS                      & H                       & 16000                                                                         & 4.52$^{+1.54}_{-1.33}$                                                                          & 85.25$^{+94.06}_{-49.90}$                                                                    & 12323.43$\pm$117.11                                                                & 52.50$\pm$0.99                                                                           \\
					2MASS                      & K$_s$                   & 22000                                                                         & 5.10$^{+1.97}_{-1.66}$                                                                          & 83.86$^{+97.06}_{-50.31}$                                                                   & 9284.47$\pm$88.25                                                                  & 47.18$\pm$0.90                                                                          \\
					\emph{WISE}                       & W1 (3.4 $\mu$m)          & 34000                                                                         & 3.87$^{+0.64}_{-0.65}$                                                                          & 74.10$^{+9.42}_{-9.42}$                                                                      & 4749.84$\pm$68.87                                                                  & 42.48$\pm$1.23                                                                          \\
					\emph{Spitzer}-IRAC                    & I1 (3.6 $\mu$m)          & 36000                                                                         & 4.10$^{+1.04}_{-0.71}$                                                                          & 72.00$^{+40.28}_{-39.53}$                                                                    & 4779.20$\pm$71.69                                                                  & 42.19$\pm$1.27                                                                          \\
					\emph{Spitzer}-IRAC                    & I2 (4.5 $\mu$m)          & 45000                                                                         & 4.84$^{+0.80}_{-0.82}$                                                                          & 83.67$^{+26.64}_{-27.30}$                                                                    & 2891.50$\pm$43.37                                                                  & 42.77$\pm$1.28                                                                          \\
					\emph{WISE}                       & W2 (4.6 $\mu$m)          & 46000                                                                         & 4.04$^{+0.68}_{-0.68}$                                                                          & 72.23$^{+11.14}_{-11.14}$                                                                    & 2600.69$\pm$44.21                                                                  & 40.76$\pm$1.38                                                                          \\
					\emph{Spitzer}-IRAC                    & I3 (5.8 $\mu$m)          & 58000                                                                         & 4.04$^{+1.39}_{-1.14}$                                                                          & 124.79$^{+61.10}_{-56.30}$                                                                   & 2549.24$\pm$53.24                                                                  & 44.20$\pm$1.34                                                                         \\
					
					\bottomrule
			\end{tabular}}
			\begin{tablenotes}
				\scriptsize
				\item \textbf{Notes.} --- The columns from left to right represent the telescope name, the filter name, the central wavelength for the filter, the \sersic{} index, the effective radius, the flux density, and the flux ratio.
			\end{tablenotes}
		\end{table}
	\end{center}
	
	\begin{center}
		\begin{table}[h]
			\centering
			\scriptsize
			\caption{Multiwavelength Statistics of the Disk as a Result of the Decomposition}
			\label{table_disk}
			\setlength{\tabcolsep}{0.1mm}{
				\begin{tabular}{ccrrrrr}
					\toprule{}
					\multirow{1}{*}{Telescope} & \multirow{1}{*}{Filter} & \multirow{1}{*}{\begin{tabular}[c]{@{}c@{}}Wavelength\\ \\ {(}\AA{)}\end{tabular}} & \multirow{1}{*}{\begin{tabular}[c]{@{}c@{}}\sersic{} index \\ ($n$)\end{tabular}} & \multirow{1}{*}{\begin{tabular}[c]{@{}c@{}}Effective radius\\ ($R_e$) \\ {(}arcsec{)}\end{tabular}} & \multirow{1}{*}{\begin{tabular}[c]{@{}c@{}}Flux denisity\\ \\ {(}mJy{)}\end{tabular}} & \multirow{1}{*}{\begin{tabular}[c]{@{}c@{}}Flux ratio\\ \\ {(}\%{)}\end{tabular}} \\
					&                         &                                                                               &                                                                                                  &                                                                                              &                                                                                    &                                                                                \\
					&                         &                                                                               &                                                                                                  &                                                                                              &                                                                                    &                                                                                \\ 					
					\midrule{}
					\emph{GALEX}                      & FUV                     & 1516                       & 0.15$^{+0.07}_{-0.03}$                               & 442.01$^{+24.62}_{-26.18}$                        & 85.29$\pm$1.92                          & 60.08$\pm$2.71                               \\
					\emph{Swift}                      & UVW2                    & 1928                       & 0.29$^{+0.03}_{-0.03}$                               & 389.38$^{+21.63}_{-19.20}$                        & 176.80$\pm$2.39                         & 84.87$\pm$2.29                               \\
					\emph{GALEX}                      & NUV                     & 2267                       & 0.20$^{+0.01}_{-0.01}$                               & 415.23$^{+6.35}_{-5.65}$                          & 202.47$\pm$2.73                         & 90.21$\pm$2.44                               \\
					\emph{Swift}                      & UVW1                    & 2600                       & 0.33$^{+0.07}_{-0.05}$                               & 361.72$^{+29.48}_{-22.15}$                        & 279.64$\pm$3.78                         & 79.32$\pm$2.14                               \\
					SDSS                       & {\it u}                       & 3551                       & 0.42$^{+0.03}_{-0.03}$                               & 335.12$^{+73.75}_{-73.53}$                        & 606.29$\pm$2.44                         & 50.94$\pm$0.41                               \\
					WIYN                       & B                       & 4331                       & 0.67$^{+0.11}_{-0.12}$                               & 309.56$^{+38.59}_{-38.60}$                        & ---                                    & ---                                \\
					SDSS                       & {\it g}                       & 4686                       & 0.94$^{+0.21}_{-0.20}$                               & 296.86$^{+35.59}_{-35.26}$                        & 2735.09$\pm$10.94                       & 59.62$\pm$0.48                               \\
					WIYN                       & V                       & 5500                       & 0.97$^{+0.22}_{-0.16}$                               & 272.53$^{+21.10}_{-20.50}$                        & ---                                    & ---                                \\
					SDSS                       & {\it r}                       & 6166                       & 1.10$^{+0.48}_{-0.36}$                               & 278.13$^{+68.45}_{-68.63}$                        & 5244.50$\pm$20.98                       & 57.99$\pm$0.46                               \\
					WIYN                       & R                       & 6425                       & 0.97$^{+0.17}_{-0.17}$                               & 281.13$^{+34.88}_{-34.90}$                        & ---                                    & ---                                \\
					SDSS                       & {\it i}                       & 7480                       & 1.18$^{+0.42}_{-0.24}$                               & 273.86$^{+68.45}_{-68.63}$                        & 7532.75$\pm$26.37                       & 57.84$\pm$0.40                               \\
					SDSS                       & {\it z}                       & 8932                       & 1.21$^{+0.40}_{-0.37}$                               & 244.20$^{+62.18}_{-62.35}$                        & 8918.78$\pm$35.69                       & 53.96$\pm$0.43                              \\
					2MASS                      & J                       & 12000                      & 1.05$^{+0.05}_{-0.05}$                               & 194.65$^{+34.94}_{-34.94}$                        & 10512.44$\pm$89.37                      & 51.71$\pm$0.88                               \\
					2MASS                      & H                       & 16000                      & 0.98$^{+0.27}_{-0.30}$                               & 191.40$^{+44.45}_{-42.39}$                        & 11252.15$\pm$106.94                     & 47.94$\pm$0.91                               \\
					2MASS                      & K$_s$                   & 22000                      & 1.10$^{+0.29}_{-0.30}$                               & 199.40$^{+50.62}_{-50.37}$                        & 10416.70$\pm$99.05                      & 52.93$\pm$1.00                               \\
					\emph{WISE}                       & W1 (3.4 $\mu$m)          & 34000                      & 1.15$^{+0.10}_{-0.10}$                               & 258.56$^{+2.72}_{-2.73}$                          & 6279.24$\pm$91.05                       & 56.15$\pm$1.62                               \\
					\emph{Spitzer}-IRAC                    & I1 (3.6 $\mu$m)          & 36000                      & 1.09$^{+0.04}_{-0.08}$                               & 247.47$^{+4.15}_{-9.66}$                          & 6349.69$\pm$95.25                       & 56.06$\pm$1.68                               \\
					\emph{Spitzer}-IRAC                    & I2 (4.5 $\mu$m)          & 45000                      & 1.13$^{+0.01}_{-0.09}$                               & 236.58$^{+10.82}_{-15.44}$                         & 3773.05$\pm$56.60                       & 55.81$\pm$1.67                               \\
					\emph{WISE}                       & W2 (4.6 $\mu$m)          & 46000                      & 1.18$^{+0.12}_{-0.12}$                               & 266.68$^{+4.18}_{-4.17}$                          & 3696.27$\pm$62.84                       & 57.93$\pm$1.97                               \\
					\emph{Spitzer}-IRAC                    & I3 (5.8 $\mu$m)          & 58000                      & 0.50$^{+0.41}_{-0.40}$                               & 297.62$^{+84.12}_{-84.60}$                        & 4296.41$\pm$64.45                       & 53.51$\pm$1.63                              \\
					\bottomrule
			\end{tabular}}
			\begin{tablenotes}
				\scriptsize
				\item \textbf{Notes.} --- This table is in the same format as Table \ref{table_bulge}
			\end{tablenotes}
		\end{table}
	\end{center}
	
 \clearpage

	\clearpage
	
	\section{\textbf{Products of the Multiwavelength Bulge-Disk Decomposition for M81 with Equalized S/N}}\label{Sec_Appendix_B}
	
	This appendix presents products of the multiwavelength bulge-disk decomposition for M81 in the images with S/N reduced to an equal level, corresponding to the description in Section \ref{Sec_Disc_SNR}.  Morphological and flux parameters for the bulge and the disk are listed in Tables \ref{table_bulge_lowsnr} and \ref{table_disk_lowsnr}, respectively.
		
	\setcounter{table}{0}
	\begin{center}
		\begin{table}[h]
			\centering
			\scriptsize
			\caption{Multiwavelength Statistics of the Bulge as a Result of the Decomposition with the Equalized S/N}
			\label{table_bulge_lowsnr}
			\setlength{\tabcolsep}{0.1mm}{
				\begin{tabular}{ccrrrrr}
					\toprule{}
					\multirow{1}{*}{Telescope} & \multirow{1}{*}{Filter} & \multirow{1}{*}{\begin{tabular}[c]{@{}c@{}}Wavelength\\ \\ {(}\AA{)}\end{tabular}} & \multirow{1}{*}{\begin{tabular}[c]{@{}c@{}}\sersic{} index \\ ($n$)\end{tabular}} & \multirow{1}{*}{\begin{tabular}[c]{@{}c@{}}Effective radius\\ ($R_e$) \\ {(}arcsec{)}\end{tabular}} & \multirow{1}{*}{\begin{tabular}[c]{@{}c@{}}Flux denisity\\ \\ {(}mJy{)}\end{tabular}} & \multirow{1}{*}{\begin{tabular}[c]{@{}c@{}}Flux ratio\\ \\ {(}\%{)}\end{tabular}} \\
					&                         &                                                                               &                                                                                                  &                                                                                              &                                                                                    &                                                                                \\
					&                         &                                                                               &                                                                                                  &                                                                                              &                                                                                    &                                                                                \\ 						\midrule{}
					\emph{GALEX}                      & FUV                     & 1516                                                                          & 0.97$^{+0.21}_{-0.21}$                                                                          & 10.08$^{+0.96}_{-0.96}$                                                                      & 0.68$\pm$0.02                                                                      & 0.48$\pm$0.02                                                                           \\
					\emph{Swift}                      & UVW2                    & 1928                                                                          & 1.32$^{+0.43}_{-0.43}$                                                                          & 25.24$^{+10.36}_{-10.36}$                                                                      & 7.27$\pm$0.10                                                                      & 3.44$\pm$0.09                                                                           \\
					\emph{GALEX}                      & NUV                     & 2267                                                                          & 1.04$^{+0.30}_{-0.30}$                                                                          & 24.23$^{+2.34}_{-2.34}$                                                                      & 5.65$\pm$0.08                                                                      & 2.48$\pm$0.06                                                                           \\
					\emph{Swift}                      & UVW1                    & 2600                                                                          & 1.93$^{+0.23}_{-0.23}$                                                                          & 50.22$^{+4.39}_{-4.39}$                                                                      & 42.68$\pm$0.58                                                                     & 12.03$\pm$0.32                                                                          \\
					SDSS                       & {\it u}                       & 3551                                                                          & 3.30$^{+1.04}_{-1.04}$                                                                          & 109.13$^{+75.69}_{-75.69}$                                                                    & 467.86$\pm$1.89                                                                    & 39.34$\pm$0.30                                                                          \\
					SDSS                       & {\it g}                       & 4686                                                                          & 3.86$^{+0.48}_{-0.48}$                                                                          & 82.32$^{+15.72}_{-15.72}$                                                                    & 1787.92$\pm$7.15                                                                   & 39.02$\pm$0.31                                                                          \\
					SDSS                       & {\it r}                       & 6166                                                                          & 4.55$^{+1.14}_{-1.14}$                                                                          & 97.24$^{+45.22}_{-45.22}$                                                                    & 3833.47$\pm$15.33                                                                  & 42.27$\pm$0.34                                                                          \\
					SDSS                       & {\it i}                       & 7480                                                                          & 4.40$^{+1.38}_{-1.38}$                                                                          & 99.06$^{+53.63}_{-53.63}$                                                                    & 5812.92$\pm$20.35                                                                  & 45.17$\pm$0.32                                                                          \\
					SDSS                       & {\it z}                       & 8932                                                                          & 4.85$^{+1.35}_{-1.35}$                                                                          & 99.32$^{+61.31}_{-61.31}$                                                                    & 7390.34$\pm$29.58                                                                  & 44.74$\pm$0.36                                                                          \\
					2MASS                      & J                       & 12000                                                                         & 4.87$^{+1.35}_{-1.35}$                                                                          & 87.38$^{+49.49}_{-49.49}$                                                                    & 9897.42$\pm$84.45                                                                  & 48.32$\pm$0.82                                                                          \\
					2MASS                      & H                       & 16000                                                                         & 4.49$^{+1.54}_{-1.54}$                                                                          & 84.45$^{+94.06}_{-94.06}$                                                                    & 12294.59$\pm$117.84                                                                & 52.52$\pm$1.00                                                                           \\
					2MASS                      & K$_s$                   & 22000                                                                         & 5.02$^{+1.97}_{-1.97}$                                                                          & 81.25$^{+97.06}_{-97.06}$                                                                   & 9157.23$\pm$87.04                                                                  & 46.34$\pm$0.88                                                                          \\
					\emph{WISE}                       & W1 (3.4 $\mu$m)          & 34000                                                                         & 4.21$^{+1.33}_{-1.33}$                                                                          & 79.12$^{+9.42}_{-9.42}$                                                                      & 4798.42$\pm$69.58                                                                  & 42.56$\pm$1.23                                                                          \\
					\emph{Spitzer}-IRAC                    & I1 (3.6 $\mu$m)          & 36000                                                                         & 4.91$^{+1.04}_{-1.04}$                                                                          & 85.03$^{+40.28}_{-40.28}$                                                                    & 4848.63$\pm$72.73                                                                  & 43.61$\pm$1.32                                                                          \\
					\emph{Spitzer}-IRAC                    & I2 (4.5 $\mu$m)          & 45 000                                                                         & 5.25$^{+2.49}_{-2.49}$                                                                          & 95.50$^{+71.90}_{-71.90}$                                                                    & 2964.83$\pm$44.47                                                                  & 43.97$\pm$1.32                                                                          \\
					\emph{WISE}                       & W2 (4.6 $\mu$m)          & 46000                                                                         & 4.78$^{+0.83}_{-0.83}$                                                                          & 83.78$^{+22.30}_{-22.30}$                                                                    & 2735.86$\pm$46.51                                                                  & 43.68$\pm$1.50                                                                          \\
					\bottomrule
			\end{tabular}}
			\begin{tablenotes}
				\scriptsize
				\item \textbf{Notes.} --- This table is in the same format as Table \ref{table_bulge}.
			\end{tablenotes}
		\end{table}
	\end{center}
	
	\begin{center}
		\begin{table}[h]
			\centering
			\scriptsize
			\caption{Multiwavelength Statistics of the Disk as the Result of the Decomposition with the Equalized S/N}
			\label{table_disk_lowsnr}
			\setlength{\tabcolsep}{0.1mm}{
				\begin{tabular}{ccrrrrr}
					\toprule{}
					\multirow{1}{*}{Telescope} & \multirow{1}{*}{Filter} & \multirow{1}{*}{\begin{tabular}[c]{@{}c@{}}Wavelength\\ \\ {(}\AA{)}\end{tabular}} & \multirow{1}{*}{\begin{tabular}[c]{@{}c@{}}\sersic{} index \\ ($n$)\end{tabular}} & \multirow{1}{*}{\begin{tabular}[c]{@{}c@{}}Effective radius\\ ($R_e$) \\ {(}arcsec{)}\end{tabular}} & \multirow{1}{*}{\begin{tabular}[c]{@{}c@{}}Flux denisity\\ \\ {(}mJy{)}\end{tabular}} & \multirow{1}{*}{\begin{tabular}[c]{@{}c@{}}Flux ratio\\ \\ {(}\%{)}\end{tabular}} \\
					&                         &                                                                               &                                                                                                  &                                                                                              &                                                                                    &                                                                                \\
					&                         &                                                                               &                                                                                                  &                                                                                              &                                                                                    &                                                                                \\ 					
					\midrule{}
					\emph{GALEX}                      & FUV                     & 1516                       & 0.15$^{+0.07}_{-0.07}$                               & 438.13$^{+26.62}_{-26.62}$                        & 104.95$\pm$2.36                          & 74.01$\pm$2.91                               \\
					\emph{Swift}                      & UVW2                    & 1928                       & 0.23$^{+0.11}_{-0.11}$                               & 390.81$^{+37.82}_{-37.82}$                        & 186.68$\pm$2.52                         & 88.43$\pm$2.30                               \\
					\emph{GALEX}                      & NUV                     & 2267                       & 0.19$^{+0.01}_{-0.01}$                               & 415.40$^{+6.35}_{-6.35}$                          & 207.15$\pm$2.80                         & 90.73$\pm$2.37                               \\
					\emph{Swift}                      & UVW1                    & 2600                       & 0.29$^{+0.07}_{-0.07}$                               & 354.56$^{+67.75}_{-67.75}$                        & 289.71$\pm$3.91                         & 81.64$\pm$2.14                               \\
					SDSS                       & {\it u}                       & 3551                       & 0.42$^{+0.27}_{-0.27}$                               & 354.56$^{+67.95}_{-67.95}$                        & 609.79$\pm$2.46                         & 51.27$\pm$0.39                               \\
					SDSS                       & {\it g}                       & 4686                       & 1.08$^{+0.21}_{-0.21}$                               & 293.42$^{+41.39}_{-41.39}$                        & 2750.00$\pm$11.10                       & 60.02$\pm$0.48                               \\
					SDSS                       & {\it r}                       & 6166                       & 1.20$^{+0.48}_{-0.48}$                               & 270.39$^{+68.63}_{-68.63}$                        & 5166.37$\pm$20.99                       & 56.96$\pm$0.45                               \\
					SDSS                       & {\it i}                       & 7480                       & 1.21$^{+0.42}_{-0.42}$                               & 262.25$^{+68.63}_{-68.63}$                        & 7223.63$\pm$26.38                       & 56.13$\pm$0.40                               \\
					SDSS                       & {\it z}                       & 8932                       & 1.21$^{+0.40}_{-0.40}$                               & 243.73$^{+63.77}_{-63.77}$                        & 9008.18$\pm$36.05                       & 54.53$\pm$0.43                              \\
					2MASS                      & J                       & 12000                      & 1.04$^{+0.14}_{-0.14}$                               & 195.68$^{+34.94}_{-34.94}$                        & 10591.24$\pm$90.04                      & 51.71$\pm$0.88                               \\
					2MASS                      & H                       & 16000                      & 0.97$^{+0.30}_{-0.30}$                               & 191.26$^{+44.45}_{-42.39}$                        & 11268.25$\pm$107.09                     & 48.13$\pm$0.92                               \\
					2MASS                      & K$_s$                   & 22000                      & 1.10$^{+0.33}_{-0.30}$                               & 200.45$^{+39.08}_{-39.08}$                        & 10547.83$\pm$100.24                      & 53.88$\pm$1.01                               \\
					\emph{WISE}                       & W1 (3.4 $\mu$m)          & 34000                      & 1.20$^{+0.12}_{-0.12}$                               & 252.75$^{+10.16}_{-10.16}$                          & 6306.34$\pm$91.44                       & 55.94$\pm$1.61                               \\
					\emph{Spitzer}-IRAC                    & I1 (3.6 $\mu$m)          & 36000                      & 1.19$^{+0.08}_{-0.08}$                               & 239.46$^{+11.92}_{-11.92}$                          & 6402.18$\pm$96.03                       & 57.58$\pm$1.74                               \\
					\emph{Spitzer}-IRAC                    & I2 (4.5 $\mu$m)          & 45000                      & 1.13$^{+0.31}_{-0.31}$                               & 232.40$^{+15.16}_{-15.16}$                         & 3749.87$\pm$56.75                       & 55.61$\pm$1.67                               \\
					\emph{WISE}                       & W2 (4.6 $\mu$m)          & 46000                      & 1.20$^{+0.12}_{-0.12}$                               & 258.38$^{+9.97}_{-9.97}$                          & 3596.17$\pm$62.88                       & 57.42$\pm$1.97                               \\
					\bottomrule
			\end{tabular}}
			\begin{tablenotes}
				\scriptsize
				\item \textbf{Notes.} --- This table is in the same format as Table \ref{table_bulge}.
			\end{tablenotes}
		\end{table}
	\end{center}

\end{document}